\documentclass[usenatbib]{mnras}
\usepackage{graphicx}
\usepackage{amsmath}
\usepackage{amssymb}
\usepackage{bm}
\usepackage{hyperref}
\usepackage{natbib}
\usepackage{color}
\usepackage{xcolor}
\pagecolor{white}
\usepackage{multirow}
\usepackage[normalem]{ulem}

\newcommand{\beq}{\begin{equation}}
\newcommand{\eeq}{\end{equation}}

\newcommand{\benum}{\begin{enumerate}}
\newcommand{\eenum}{\end{enumerate}}

\def\apj{ApJ}

\def\apj{ApJ}

\def\pre{Physical Review E}

\def\prl{Phys. Rev. Lett.}

\newcommand{\abra}[1]{\left\langle{#1}\right\rangle}

\newcommand{\bmB}{{\bm B}}
\newcommand{\bmb}{{\bm b}}
\newcommand{\bmU}{{\bm U}}
\def\bmu{\bm{u}}
\newcommand{\bmJ}{{\bm J}}

\newcommand{\bmk}{{\bm{k}}}

\newcommand{\bmx}{{\bm{x}}}

\newcommand{\emf}{\mathcal{E}}

\newcommand{\del}{{\bm\nabla}}
\newcommand{\nuM}{{\nu_\text{M}}}

\newcommand{\taucor}{{\tau_\text{cor}}}

\newcommand{\Sh}{\text{Sh}}

\newcommand{\ReM}{\text{Rm}}

\newcommand{\Co}{\text{Co}}
\newcommand{\ReN}{\text{Re}}
\newcommand{\kf}{k_\text{f}}
\newcommand{\Ro}{\text{Ro}}
\newcommand{\urms}{u_\text{rms}}
\newcommand{\dl}{\partial^l}
\newcommand{\ddl}{{\partial^{l}}^2}
\newcommand{\bml}{\bm l}

\newcommand{\knu}{k_\nu}
\newcommand{\qspec}{q_\text{s}}
\newcommand{\sgn}[1]{\text{sgn}\left\{{#1}\right\}}
\newcommand{\PrM}{\text{Pm}}

\begin{document}

\title[On the shear current effect]
{On the shear-current effect: toward understanding why theories and simulations have mutually and separately conflicted}
\author[Zhou \& Blackman]
{Hongzhe Zhou$^{1}$%
\thanks{Email address for correspondence: hongzhe.zhou@su.se},
Eric G. Blackman$^{2,3}$%
\thanks{Email address for correspondence: blackman@pas.rochester.edu}\\
$^1$
Nordita, KTH Royal Institute of Technology and Stockholm University,
Hannes Alfv\'ens v\"ag 12, SE-106 91 Stockholm, Sweden\\
$^2$ Department of Physics and Astronomy, University of Rochester, Rochester, NY, 14627, USA\\
$^3$ Laboratory for Laser Energetics, University of Rochester, Rochester NY, 14623, USA\\
}

\date{\today}
\maketitle

\begin{abstract}
The shear-current effect (SCE) of mean-field dynamo theory refers to the combination of a shear flow and a turbulent coefficient $\beta_{21}$ with a favorable negative sign for exponential mean-field growth, rather than positive for diffusion. 
There have been long standing disagreements among theoretical calculations and comparisons of theory with numerical experiments as to the sign of kinetic ($\beta^u_{21}$) and magnetic ($\beta^b_{21}$) contributions.
To resolve these discrepancies, we combine an analytical approach with simulations, and show that unlike $\beta^b_{21}$, the kinetic SCE $\beta^u_{21}$ has a strong dependence on the kinetic energy spectral index and can transit from positive to negative values at $\mathcal{O}(10)$ Reynolds numbers if the spectrum is not too steep. Conversely, $\beta^b_{21}$ is always negative regardless of the spectral index and Reynolds numbers.
For very steep energy spectra, the positive $\beta^u_{21}$ can dominate even at energy equipartition $\urms\simeq b_\text{rms}$, resulting in a positive total $\beta_{21}$ even though $\beta^b_{21}<0$.
Our findings bridge the gap between the seemingly contradictory results from 
the second-order-correlation approximation (SOCA) versus the spectral-$\tau$ closure (STC), for which opposite signs for $\beta^u_{21}$ have been reported, with the same sign for $\beta^b_{21}<0$.
The results also offer an explanation for the simulations that find 
$\beta^u_{21}>0$ and an inconclusive overall sign of $\beta_{21}$ for $\mathcal{O}(10)$ Reynolds numbers.
The transient behavior of $\beta^u_{21}$ is demonstrated using the kinematic test-field method.
We compute dynamo growth rates for cases with or without rotation, and discuss opportunities for further work. 
\end{abstract}

\section{Introduction}
\subsection{Background}
Dynamos that amplify and sustain magnetic fields are believed to operate in a wide range of astrophysical systems.
Depending on whether the spatio-temporal scale of the amplified magnetic fields is smaller or larger than the energy-dominant scale of the turbulent flow, dynamos can be classified into ``small scale'' and ``large scale'' types.
Mean-field dynamo theory is a commonly adopted framework for studying large scale dynamos, where statistical properties of the turbulence is important.
Widely employed in different approaches, a non-zero average kinetic helicity of the hosting turbulent flow greatly helps large scale magnetic field amplification--the so-called $\alpha$ effect
 \citep[e.g. ][]{Parker1955,Steenbeck1966,Pouquet1976,Moffatt1978,Parker1979,Blackman2002prl,BrandenburgSubramanian2005}.

However, it is less clear whether a non-kinetically helical turbulent flow might also generate a large-scale field.
This might be important for systems with weak density stratification such as midplanes of 
accretion discs, or possibly even planetary cores, if inertial waves are an insufficient source of kinetic helicity \citep{Moffatt1970,Olson1981,DavidsonRanjan2018}.

The shear-current effect (SCE) is one potential non-kinetically helical large-scale dynamo.
In a mean shear flow (for instance $\bm U^\text{shear}=-Sx\hat{\bm y}$ with $S>0$), the turbulent diffusion tensor for the mean magnetic field becomes anisotropic, and in particular, its $yx$-component may become negative \citep{RogachevskiiKleeorin2003, RogachevskiiKleeorin2004}. 
Then there exists a growing mode for the mean magnetic field, even without the kinetic $\alpha$ effect.
But the SCE has been controversial, and as we shall discuss in detail, interpretations from theoretical calculations using different closures have disagreed, as have theory and some numerical simulations.
Whether or not the respective contributions from the turbulent velocity and magnetic fields to the turbulent diffusion tensor have the SCE-preferred sign, when, and which dominates
are all not fully agreed upon. The answer has varied among folks using different closure approximations for the high-order turbulent correlations.
Those employing a spectral-$\tau$ closure \citep[STC;][]{RogachevskiiKleeorin2003,RogachevskiiKleeorin2004} or its minimal-$\tau$ variation \citep{Pipin2008} found that the diffusivity tensor has the SCE favorable signs for both kinetic and the magnetic parts.
In contrast, the kinetic contribution was later found to have a SCE-incompatible sign by second-order-correlation approximation (SOCA) and quasi-linear calculations
\citep{RaedlerStepanov2006, Ruediger2006,SridharSubramanian2009,SridharSingh2010,SinghSridhar2011}.

\cite{Squire2015pre} performed a SOCA calculation of the magnetic SCE, and agreed with the STC calculation that the magnetic SCE exists with favorable sign, arguing that it could dominate in the presence of a strong turbulent magnetic field produced by a small-scale dynamo.
\cite{Squire2016jpp} found that the pressure gradient ($\del p$) term in the Navier-Stokes equation was
essential for the favorable magnetic SCE contribution, although this was challenged by \cite{Kapyla2020} using SOCA [although in the ideal magnetohydrodynamics (MHD) limit], arguing that the magnetic contribution could survive even without the $\del p$ term, but with the wrong sign.

The first numerical evidence of a shear dynamo, not necessarily driven by SCE, included \cite{Brandenburg2005distributedDynamo} and \cite{Yousef2008prl,Yousef2008AN}, where a large-scale shear flow was superimposed upon non-helical forcing, and magnetic field amplification above the forcing scale was observed.
\cite{HughesProctor2009} studied the combination of a shear flow and rotating convection, and found that shear promotes a large-scale dynamo which would otherwise be subcritical.
To identify the dynamo driver, the test-field method (TFM) was commonly employed. Along with the ``main run'', a ``test-field run'' is performed in parallel, whereby the evolution of some known dynamically weak test field is measured and the turbulent transport coefficients inferred.
Kinematic \citep{Brandenburg2008magdiff,SinghJingade2015}, quasi-kinematic and non-linear \citep{Kapyla2020,Kapyla2021CTFM} TFMs all disfavored the SCE and revealed positive values for $\beta_{21}$ in both kinetically forced and kinetic-magnetically forced systems.
These authors argued that the mean-field amplification was more likely the result of the stochastic $\alpha$ effect \citep{VishniacBrandenburg1997,Heinemann2011,Mitra2012mnras,Richardson2012mnras,Newton2012phpl,SridharSingh2014mnras,Singh2016jfm,Jingade2018jpp}.

An alternative approach to obtain the turbulent transport coefficients from simulations is the projection method \citep{BrandenburgSokoloff2002,Squire2015prl,Squire2015apj,Squire2016jpp,Shi2016}.
Then a negative $\beta_{21}$ for kinetically forced rotating shearing turbulence, as well as for magnetically forced non-rotating shearing turbulence, is obtained.
These results are in agreement with the SOCA calculation \citep{Squire2015pre},
although the validity of setting some transport coefficients to zero \textit{a priori} while solving for others is unclear, an approximation commonly adopted in these works.
Recently, \cite{Wissing2021} has reported $\beta_{21}\geq0$ in magneto-rotational instability (MRI) turbulence using the projection method, conflicting \cite{Shi2016} who also worked with MRI turbulence.
Whether the origin of this inconsistency lies in MRI or the projection method warrants further study.

In short, SOCA calculations disagree with STC in the kinetic SCE but agree with simulations, whereas both SOCA and STC support the magnetic SCE but it is unclear whether direct simulations do.
The literature mentioned above is summarized in Table \ref{tab:sc} in a chronological order.

\begin{table*}
\caption{Summary of theoretical and numerical work on SCE.
Theoretical work labeled by $\dagger$ used spectral-$\tau$ or minimal-$\tau$ closure, and otherwise used SOCA or quasi-linear approximation.
Numerical work labeled by $*$ used a projection method, and otherwise used the test-field method.
}
\label{tab:sc}
\begin{tabular}{llll}
\hline
Theory & Simulation & Kinetic/ & Remarks\\
&& Magnetic SCE?&\\
\hline
$^\dagger$\cite{RogachevskiiKleeorin2003} & & Y/\ & \\
$^\dagger$\cite{RogachevskiiKleeorin2004} & & Y/Y & \\
\cite{RaedlerStepanov2006} && N/\ & \\
\cite{Ruediger2006} && N/\ & \\
& \cite{Brandenburg2008magdiff} & N/\ & $\ReN=\mathcal{O}(1),\ReM\lesssim\mathcal{O}(100)$\\
$^\dagger$\cite{Pipin2008} && Y/Y & \\
\cite{SridharSubramanian2009} && N/\ & Quasi-linear in shearing frame\\
\cite{SridharSingh2010} && N/\ & Quasi-linear in shearing frame \\
\cite{SinghSridhar2011} && N/\ & Quasi-linear in shearing frame \\
\cite{Squire2015pre} && N/Y& \\
& \cite{Squire2015apj} & N/Y & $\ReN=\ReM\simeq 5$\\
& $^*$\cite{Squire2015prl} & N/Y & $\ReN=\ReM\lesssim15$\\
& \cite{SinghJingade2015} & N/\ & $\text{min}\left\{\ReN,\ReM\right\}<1$\\
&$^*$\cite{Squire2016jpp} & N/Y & $8\ReN=\ReM\lesssim5\abra{u^2}/\abra{B^2}$\\
& \cite{Kapyla2020} & N/N & MHD burgulence, $\ReN<1$, $\ReM<15$\\
\hline
\end{tabular}
\end{table*}

Typically the SCE is discussed within the kinematic dynamo phase, 
i.e., when the mean magnetic field is dynamically weak and its backreaction on the turbulent flow can be neglected. We also work within this regime in this paper.
The nonlinear phase and saturation of the shear dynamo remains elusive.
The joint effect of the shear-enhanced small-scale dynamo and the vorticity dynamo which enhances shear further complicates the problem.
For references, see \cite{Rogachevskii2006}, \cite{Teed2016}, and \cite{Singh2017}.

\subsection{Aim and path of the paper}
We first investigate the origin of the theoretical contradiction between SOCA and STC in the SCE context.
In SOCA, nonlinear terms in the Navier-Stokes and the small-scale induction equations are dropped, as is justified at low hydro and magnetic Reynolds numbers, or at small Strouhal numbers.
In STC, the nonlinear terms are replaced by an eddy-damping term but the microscopic diffusion terms are dropped, as is justified at high Reynolds numbers.
To elucidate the difference between these two different choices, we keep both the viscous terms and the eddy-damping terms.
We then examine the sign of both the kinetic and magnetic contributions of $\beta_{21}$ at the order linear in the shear rate, and show how they each depend on the Reynolds numbers and the energy spectral indices.
We also use the kinematic TFM to validate our findings.

We then compute the full diffusivity tensor nonlinearly, but still perturbatively, in the shear rate $S$. This includes $4$ components for the kinetic contribution and $4$ for the magnetic part. We include the spatial inhomogeneity of the mean flow to the third order in $S$, or first order in $S$ in the presence of rotation, while the shear-dependence of other terms in the equations are treated exactly.
We do not solve for the anisotropic corrections to the velocity and magnetic auto-correlations, but assume that they are isotropic and nonhelical, and validate this 
assumption for slow rotation with simulations.
For incompressible turbulence, we compare the resulting diffusivity tensors in cases with or without the pressure gradient term, and with or without a Keplerian rotation, and discuss implications for shear-current dynamos in shearing boxes and astrophysical dynamos.

In Section \ref{sec:analytical} we use a modified spectral-$\tau$ approach to derive the turbulent diffusivity tensor.
In Section \ref{sec:Re_dependence} we focus on $\beta_{21}$, and show how its kinetic and magnetic contributions depend differently on the Reynolds numbers and spectral indices.
We discuss the implications of our findings for simulations in Section \ref{sec:simulation}.
In Section \ref{sec:result} we present the full turbulent diffusivity tensor and the corresponding dynamo growth rates.
We conclude in Section \ref{sec:conclusion}.

\section{Calculation of the diffusivity tensor}
\label{sec:analytical}
We consider a Cartesian geometry with periodic boundaries and $xy$-planar averaged mean fields.
See Section \ref{sec:bc} for a discussion of the boundary conditions.
These choices can only be justified at low shear rates.
We assume a sufficiently large scale seperation between mean and fluctuating fields, and thus an equivalence between the planar average and an ensemble average \citep{Hoyng1988,Zhou2018}, both denoted by angle brackets.
The total velocity field is $\bmU^\text{tot}=-Sx\hat{\bm y}+\bm U+\bmu$ with a constant shear rate $S>0$, where $\bm U$ is a planar-averaged mean flow that may arise due to a vorticity dynamo \citep{Elperin2003,Kapyla2009}, and the fluctuating field $\bmu$ has a zero mean.
We decompose the total magnetic field as $\bmB^\text{tot}=\bmB+\bmb$, where $\bmB$ is the mean field and $\bmb$ is the fluctuation.
We also assume incompressibility for $\bmu$, and statistical homogeneity of $\bmu$ and $\bmb$.
Because of the planar average, we have $\partial_x\bmB=\partial_y\bmB=\bm 0$, and the divergence-free condition implies that $B_z$ is a constant, which we choose to be $0$.
The Navier-Stokes and induction equations are then
\begin{align}
\partial_t\bm u=&
-\bm u\cdot\bm\nabla\bm u
+Sx\partial_y \bm u+S u_x \hat{\bm y}
-\bm\nabla p-\frac{2}{q}S\hat{\bm z}\times\bmu
\notag\\
&+\bmB\cdot\del\bmb+\bmb\cdot\del\bmB
+\bmb\cdot\del\bmb
\notag\\
&+\abra{\bmu\cdot\del\bmu}
-\abra{\bmb\cdot\del\bmb}
+\nu\nabla^2\bm u
+\bm f,
\label{eqn:NS}\\
\partial_t\bmb=&
-Sb_x\hat{\bm y}+Sx\partial_y\bmb
+\bmB\cdot\del\bmu-\bmu\cdot\del\bmB
\notag\\
&+\del\times\left(\bmu\times\bmb-\bm\emf\right)
\notag\\
&+\bmb\cdot\del\bm U-\bm U\cdot\del\bmb
+\nuM\nabla^2\bmb,
\label{eqn:induc_b}\\
\partial_t\bmB=&\del\times\bm\emf-SB_x\hat{\bm y}+\nuM\nabla^2\bmB
+\bmB\cdot\del\bm U-\bm U\cdot\del\bmB,
\end{align}
and the turbulent electromotive force (EMF) is $\bm\emf=\int dxdy\ \bmu\times\bmb\simeq\abra{\bmu\times\bmb}$.
Here, magnetic fields are written in velocity units, $p$ is the total pressure, $\bm f$ is an isotropic non-helical kinetic forcing, and $\nu$ and $\nuM$ are microscopic viscosity and diffusivity, respectively.
We have also included the Coriolis force using $\bm\Omega=S\hat{\bm z}/q$, and $q=3/2$ corresponds to a Keplerian rotation.
Note that the pressure term includes both the thermal and the magnetic pressure, because we have rewritten the Lorentz force as
\beq
\left(\del\times\bmB^\text{tot}\right)\times\bmB^\text{tot}
=-\frac{1}{2}\del {B^\text{tot}}^2+\bmB^\text{tot}\cdot\del\bmB^\text{tot}.
\eeq
Consequently, the pressure gradient term will still be present even in the MHD burgulence case where the thermal pressure is dropped.

In what follows, we shall neglect the 
magnetic 
dynamo effect from the mean flow $\bm U$, and take $\bm U=\bm 0$ with or without rotation.
While in rotating shearing flows the Rayleigh stability criterion implies the absence of a vorticity dynamo when $0<q<2$, taking $\bm U=\bm 0$ restricts the theory here to pure shear flows. 
Including the vorticity dynamo effect requires including a turbulent ponderomotive force, which we do not investigate in this work.
In our numerical investigations here, we always subtract away $\bm U$ in the shearing box by hand, thereby removing any possible magnetic dynamo action brought by it.

The EMF can be expanded in spatial gradients of $\bmB$ to close the equations. In doing so, contributions are commonly divided into several terms according to their symmetry properties. For example,
\beq
\emf_i=\text{terms linear in }\bmB
-\eta J_i+\epsilon_{ijk}\Delta_j J_k+
\kappa_{ijk}\Lambda_{jk}+\cdots,
\label{eqn:emf_decomposition}
\eeq
where $\bmJ=\del\times\bmB$, $\Lambda_{ij}=\partial_i B_j$, $\eta$ is a scalar, $\Delta_i$ is a pseudo-vector, and $\kappa$ is a tensor symmetric in $j\leftrightarrow k$.
The $\bm\Delta$ term on the right is associated with the R\"adler effect when there is a global rotation.
In our case, since $\partial_x\bmB=\partial_y\bmB=\bm 0$, the $\Lambda_{ij}$ tensor can be expressed solely in terms of $\bmJ$, and in fact $\bm J=(-\Lambda_{32},\Lambda_{31},0)$\footnote{%
We use $(1,2,3)$ and $(x,y,z)$ interchangeably in the subscripts.%
}.
We can therefore collectively write
\beq
\emf_i=\text{terms linear in }\bmB-\beta_{ij}J_j,
\eeq
and express $\beta_{ij}$ in terms of the correlation functions of $\bmu$ and $\bmb$.
The SCE is associated with a negative $\beta_{21}$.

\subsection{Equations of turbulence correlations}
The calculations will be performed up to the first spatial derivative of the mean magnetic field, i.e., linear in $\bm\Lambda$ or $\bm J$.
The pressure term is eliminated for an incompressible velocity field using the projection operator $\hat P_{ij}=\delta_{ij}-\partial_i\partial_j/\partial^2$ in Equation (\ref{eqn:NS}).
We obtain
\begin{align}
\partial_t u_i
=&B_m \partial_m b_i+
\left(2\hat P_{ij}-\delta_{ij}\right) b_m \Lambda_{mj}
\notag\\
&+S\hat P_{ij}(x\partial_2 u_j)
+S\hat P_{i2}u_1
-\frac{2S}{q}\hat P_{im} \epsilon_{m3n}u_n
\notag\\
&+\nu\nabla^2 u_i+T^u_i,
\label{eqn:dudt}\\
\partial_t b_i
=&B_m\partial_m u_i
-u_m \Lambda_{mi}
+Sx\partial_2b_i
-S\delta_{i2} b_1 
\notag\\
&+\nuM\nabla^2 b_i+T^b_i,
\label{eqn:dbdt}
\end{align}
where
\begin{align}
T^u_i=&\hat P_{ij}\left\{
\bmb\cdot\del\bmb
-\bmu\cdot\del\bmu
+\abra{\bmu\cdot\del\bmu-\bmb\cdot\del\bmb}
+\bm f
\right\}_j,\\
{\bm T}^b=&\del\times\left(\bmu\times\bmb-\bm\emf\right).
\end{align}

Denote
\begin{align}
K_{ij}(\bm l)=\int d^3x\ \abra{u_i(\bmx) u_j(\bmx+\bm l)},
\label{eqn:Kij}\\
M_{ij}(\bm l)=\int d^3x\ \abra{b_i(\bmx) b_j(\bmx+\bm l)},\\
C_{ij}(\bm l)=\int d^3x\ \abra{u_i(\bmx) b_j(\bmx+\bm l)},
\end{align}
where $\bml$ is a displacement vector.
The time evolution of $C_{ij}$ can be derived from Equations (\ref{eqn:dudt}) and (\ref{eqn:dbdt}):
\beq
\partial_t C_{ij}=Sl_1\dl_2 C_{ij}
+T^B_{ij}+T^\Lambda_{ij}+T^S_{ij}+D_{ij}+Q_{ij},
\label{eqn:Cij}
\eeq
where $\dl_i\equiv\partial/\partial l_i$, and
\begin{align}
T^B_{ij}=&\dl_m \int d^3x\ \left[
B_m(\bmx+\bm l)\abra{u_i(\bmx)u_j(\bmx+\bm l)}\right.
\notag\\
&\left.-B_m(\bmx)\abra{b_i(\bmx)b_j(\bmx+\bm l)}
\right],
\label{eqn:TBij}\\
T^\Lambda_{ij}=&-K_{im}\Lambda_{mj}
+\left(\delta_{im}-2\frac{\dl_i\dl_m}{\ddl}\right)M_{lj}\Lambda_{lm},\\
T^S_{ij}=&S\left(\delta_{i2}-2\frac{\dl_i\dl_2}{\ddl}\right)C_{1j}
-S\delta_{j2}C_{i1}\notag\\
&-\frac{2S}{q}\left(\delta_{im}-\frac{\dl_i\dl_m}{\ddl}\right)
\epsilon_{m3n}C_{nj},
\label{eqn:TSij}\\
D_{ij}=&(\nu+\nuM)\dl_m\dl_m C_{ij},\\
Q_{ij}=&\int d^3x\ T^u_i(\bmx)b_j(\bmx+\bm l)+u_i(\bmx)T^b_j(\bmx+\bm l).
\end{align}
Here we have omitted writing the $\bm l$-dependence for the quantities which depend only on $\bm l$.
In the derivation we have used
\begin{align}
&\int d^3x\ B_m(\bmx)\abra{\partial_m b_i(\bmx)b_j(\bmx+\bml)}\notag\\
&=-\int d^3x\ B_m(\bmx)\abra{b_i(\bmx)\partial_mb_j(\bmx+\bml)}\notag\\
&=-\int d^3x\ B_m(\bmx)\abra{b_i(\bmx)\dl_mb_j(\bmx+\bml)}\notag\\
&=-\dl_m\int d^3x\ B_m(\bmx)\abra{b_i(\bmx)b_j(\bmx+\bml)},
\end{align}
and similarly for other terms, assuming that the differential operator is exchangeable with the integration.
We have also taken $\Lambda_{ij}$ as a constant and pulled it out from the integrals in $T^\Lambda_{ij}$, as appropriate when we are not interested in terms of order $\del\del\bmB$ or higher.

It is more convenient to solve Equation (\ref{eqn:Cij}) in Fourier space.
We denote the Fourier transform of a field $F({\bml})$ by a tilde and define it as
\beq
\tilde F(\bmk)=\int d^3l\ F(\bml)e^{-i\bmk\cdot\bml}.
\eeq
The Fourier transforms of the turbulent correlation functions with respect to $\bml$ are just
\begin{align}
&\tilde K_{ij}(\bmk)=\abra{\tilde u_i^*(\bmk)\tilde u_j(\bmk)},\\
&\tilde M_{ij}(\bmk)=\abra{\tilde b_i^*(\bmk)\tilde b_j(\bmk)},\\
&\tilde C_{ij}(\bmk)=\abra{\tilde u_i^*(\bmk)\tilde b_j(\bmk)}.
\end{align}
Assuming that $\tilde{\bmu}(\bmk)$ and $\tilde{\bmb}(\bmk')$ only correlate at $\bmk=-\bmk'$, the EMF can be written as
\beq
\emf_i=\epsilon_{ijk}\int d^3k\ \tilde C_{jk}(\bmk).
\eeq

In Equation (\ref{eqn:Cij}), a Fourier transformation with respect to $\bml$ leads to
\beq
\partial_t\tilde C_{ij}
+Sk_2\frac{\partial}{\partial k_1}\tilde C_{ij}
-\tilde T^S_{ij}-\tilde D_{ij}-\tilde Q_{ij}
=\tilde T^\Lambda_{ij}+\tilde T^B_{ij},
\label{eqn:tCij}
\eeq
where we define $P_{ij}=\delta_{ij}-k_ik_j/k^2$,
\begin{align}
\tilde T^S_{ij}=&S\left(2P_{i2}-\delta_{i2}\right)\tilde C_{1j}
-S\delta_{j2}\tilde C_{i1}
-\frac{2S}{q}P_{im}
\epsilon_{m3n}\tilde C_{nj},
\label{eqn:tTSij}\\
\tilde T^\Lambda_{ij}=&-\tilde K_{im}\Lambda_{mj}
+(2P_{im}-\delta_{im})\tilde M_{lj}\Lambda_{lm},\\
\tilde T^B_{ij}=&-\frac{1}{2}
\Lambda_{nm}k_m\frac{\partial}{\partial k_n}
\left(\tilde K_{ij}+\tilde M_{ij}\right)\notag\\
&+\text{terms linear in } \bmB,
\label{eqn:tTBij}\\
\tilde D_{ij}=&-(\nu+\nuM)k^2\tilde C_{ij},
\end{align}
and $\tilde Q_{ij}$ is the Fourier transform of $Q_{ij}$ with respect to $\bm l$.
The Fourier transform of $T^B_{ij}$ is derived in Appendix \ref{appx:TBij}.

\subsection{The closure}
Both SOCA \citep{RaedlerStepanov2006, Ruediger2006, Squire2015pre} and STC \citep{RogachevskiiKleeorin2003,RogachevskiiKleeorin2004} closures have been applied to the SCE, and have yielded opposite signs for the kinetic contribution.

In SOCA, one drops the second-order correlations in the Navier-Stokes and induction equations for the small-scale fields, and $\bmu$ and $\bmb$ become exactly solvable.
Doing so is justified when either the Reynolds numbers or the Strouhal number is small. The correlation tensors of small-scale fields then consist of a background component given by the forcing, and correction terms perturbative in the mean quantities like shear or the mean magnetic field.
The resulting EMF is expressed in terms of two-time two-point correlations of $\bmu$ and $\bmb$.

In STC, the forcing is assumed to be weakly coupled with the small-scale magnetic fields so that $\abra{\tilde f_i^* b_j}\simeq 0$, while the sum of the viscous and the triple correlation terms are replaced by a damping term $-{\tilde C_{ij}}/{\taucor}$, and typically $\taucor$ is chosen to be of the form
\beq
\taucor=\tau \left(\frac{k}{\kf}\right)^{-\lambda},
\label{eqn:taucor}
\eeq
where $\kf$ is the forcing or energy dominant scale of the turbulence, and $\tau=1/(\urms\kf)$ is the eddy turnover time at $k=\kf$.
In STC, an effective turbulent diffusion ends up dominating microscopic diffusion which might be justified only at high Reynolds numbers.
A typical choice is $\lambda=\qspec-1$ where $\qspec$ is the spectral index of the turbulent kinetic energy \citep{Raedler2003,RogachevskiiKleeorin2003,RogachevskiiKleeorin2004, BrandenburgSubramanian2005}.

In this work we use a ``hybrid'' approach to replace the forcing and the triple correlation terms by an eddy damping term, while also keeping the dissipation terms. We can then investigate the problem at intermediate Reynolds numbers where both effects might be influential.
The eddy-damping term is a closure rather than an approximation by itself. We will see that the existence of the SCE depends entirely on the scaling of $\taucor$ [Equation (\ref{eqn:taucor})], and thus the accuracy of STC becomes crucial to prove SCE. A detailed comparison between SOCA and STC can be found in \cite{RaedlerRheinhardt2007}.

In the presence of shear or rotation, we further generalize $\taucor$ to allow for a dependence on the shear rate or the rotation rate,
\beq
\taucor=\frac{\tau(k/\kf)^{-\lambda}}{1+[1-(\hat{\bmk}\cdot\hat{\bm z})^2]\Sh},
\eeq
so that the horizontal plane cascade time is shortened because the shear flow shreds turbulent eddies, but the vertical direction is left unaffected \citep{Blackman2015}.
Here, $\Sh=S\tau=S/(\urms\kf)$ is the shear parameter. For a rotating flow without shear, we replace $\Sh$ by $3\Co/4$, with $\Co=2\Omega/(\urms\kf)$ being the Coriolis number and $S=3\Omega/2$ is the Keplerian relation.

We seek a steady-state solution for which $\partial_tC_{ij}\ll-C_{ij}/\taucor$.
Then, after applying the closure, Equation (\ref{eqn:tCij}) can be written as
\beq
{\mathcal{N}}_{ijab}\tilde C_{ab}
+Sk_2\frac{\partial}{\partial k_1}\tilde C_{ij}
=\tilde T^B_{ij}+\tilde T^\Lambda_{ij},
\label{eqn:NCT}
\eeq
where
\begin{align}
{\mathcal{N}}_{ijab}
=&S \delta_{j2}\delta_{ia}\delta_{1b}
-S (2P_{i2}-\delta_{i2})\delta_{1a}\delta_{jb}\notag\\
&+\frac{2S}{q} P_{im}\epsilon_{m3a}\delta_{jb}
+(\nu+\nuM)k^2\delta_{ia}\delta_{jb}\notag\\
&+\frac{1+[1-(\hat{\bmk}\cdot\hat{\bm z})^2]\Sh}{\tau}
\left(\frac{k}{\kf}\right)^{\lambda}
\delta_{ia}\delta_{jb}
\label{eqn:N}
\end{align}
is an algebraic tensor.
The right side of Equation (\ref{eqn:NCT}) involves no $\tilde C_{ij}$ tensor and serves as ``source terms'', for which we use homogeneous isotropic and nonhelical background turbulence,
\begin{align}
\tilde K_{ij}=&\abra{\tilde u_i^*(\bmk)u_j(\bmk)}=E^u(k)P_{ij},\\
\tilde M_{ij}=&\abra{\tilde b_i^*(\bmk)b_j(\bmk)}=E^b(k)P_{ij},
\end{align}
where, for $\kf\leq k\leq \knu$,
\begin{align}
E^u(k)=&\frac{\abra{u^2}k^{-\qspec-2}}
{8\pi\int_{\kf}^{\knu} dk\ k^{-\qspec}}
,\\
E^b(k)=&\frac{\abra{b^2}k^{-\qspec-2}}
{8\pi\int_{\kf}^{\knu} dk\ k^{-\qspec}}.
\end{align}
We have made several simplifying assumptions here:
(i) the magnetic Prandtl number $\PrM=\nu/\nuM$ is chosen to be unity, 
(ii) we have used the same dissipation wavenumber $\knu=\kf\ReN^{1/(3-\qspec)}$ for both fields, where $\ReN=\urms/(\nu\kf)$ is the Reynolds number, and
(iii) we have also used the same energy spectrum for both the velocity and magnetic fields.
These simplifying assumptions can be relaxed and 
do not change our results qualitatively, 
but simplify the present analysis.
In some previous work the first-order correction to the isotropic auto-correlation functions $\tilde K_{ij}$ and $\tilde M_{ij}$ was solved for \citep{RogachevskiiKleeorin2003,RogachevskiiKleeorin2004,RaedlerStepanov2006,Squire2015pre}. Here, we avoid this complexity and quantitatively justify our choice of isotropy for Coriolis number $\lesssim0.5$ in Appendix \ref{appx:corfunc}.

\subsection{Solving for the EMF}
Equation (\ref{eqn:NCT}) can be solved iteratively.
The zeroth-order solution is obtained by neglecting the $k$-derivative term, the first-order solution uses the zeroth-order solution in the $k$-derivative term, and so on.
The algebraic tensor $\mathcal{N}$ on the left side of Equation (\ref{eqn:NCT}) can be inverted to give the zeroth-order solution,
\beq
\tilde C_{ab}^{(0)}=\mathcal{G}_{abij}\left(
\tilde T^B_{ij}
+\tilde T^\Lambda_{ij}\right),
\eeq
where $\mathcal{G}$ is the inverse of $\mathcal{N}$ so that $\mathcal{G}_{abij}\mathcal{N}_{ijmn}=\delta_{am}\delta_{bn}$.
The $n$th-order solution of $\tilde C_{ab}$ can be found from the $(n-1)$th order by
\beq
\tilde C^{(n)}_{ab}=\tilde C_{ab}^{(0)}
-S\mathcal{G}_{abij}k_2\frac{\partial}{\partial k_1}\tilde C_{ij}^{(n-1)},
\label{eqn:induction}
\eeq
and we show in Appendix \ref{appx:Cn} that
\begin{align}
\tilde C^{(n)}_{ab}=&
\left[\tilde T^\Lambda_{cd}+\frac{1}{2}\Lambda_{nm}k_m
(\tilde K_{cd}+\tilde M_{cd})
\frac{\partial}{\partial k_n}\right]\notag\\
&\times\left[
\mathcal{G}_{abcd}
+\sum_{m=1}^n (Sk_2)^m\mathcal{H}^{(m)}_{abef}
\mathcal{G}_{efcd}\right]\notag\\
&+\text{total derivatives},
\label{eqn:Cab}
\end{align}
where
\beq
\mathcal{H}^{(1)}_{abcd}=\frac{\partial}{\partial k_1}\mathcal{G}_{abcd},\ 
\mathcal{H}^{(n)}_{abcd}=\frac{\partial}{\partial k_1}
\left[\mathcal{H}^{(n-1)}_{abef}
\mathcal{G}_{efcd}\right].
\eeq
Equation (\ref{eqn:Cab}) is independent of the closure.
In this paper we work up to third order [$\tilde C_{ab}=\tilde C_{ab}^{(3)}$] in the absence of rotation, at the first order [$\tilde C_{ab}=\tilde C_{ab}^{(1)}$] with rotating shear flow, and non-perturbatively for the pure rotation case because then $\tilde C_{ab}^{(n)}\equiv \tilde C_{ab}^{(0)}$.
In all perturbative solution cases, we have confirmed that the solutions for $\Sh\leq 0.5$ have quantitatively converged by comparing them with higher-order solutions.

The EMF is calculated through $\emf_k=\epsilon_{kab}\int d^3k\ \tilde C_{ab}(\bmk)$. Its analytical form is too cumbersome to be useful here,
so we present the results after numerical integration.
The turbulent diffusivity tensor can be split into a kinetic and a magnetic contribution, 
\beq
\emf_i=-\beta_{ij}J_j
=-\left(\beta_{ij}^u\beta^u_0
+\beta_{ij}^b\beta^b_0\right)J_j,
\label{eqn:emf_eta}
\eeq
where
\beq
\beta^u_0
=\frac{8\pi}{3}\int dk\ 
\frac{k^2 E^{u}(k)}{\nu k^2+\tau^{-1} k^{\lambda}}
\eeq
and
\beq 
\beta^b_0
=\frac{8\pi}{3}\int dk\ 
\frac{k^2 E^{b}(k)}{\nuM k^2+\tau^{-1} k^{\lambda}}
\eeq
are dimensional normalizations so that $\beta^u_{ij}=\delta_{ij}$ and $\beta^b_{ij}=0$ in the absence of shear and rotation.

\section{dependence of $\beta_{21}$ on Reynolds number and spectral indices }
\label{sec:Re_dependence}
\subsection{The non-rotating case}
We first consider the case of homogeneous non-rotating turbulence with shear, and focus on the shear-current coefficient $\beta_{21}$.
SOCA \citep{RaedlerStepanov2006, Ruediger2006, Squire2015pre} and STC \citep{RogachevskiiKleeorin2003,RogachevskiiKleeorin2004} approaches agree that the magnetic conbtribution $\beta^b_{21}$ is negative (thus favoring the SCE), but disagree on the sign of the kinetic contribution (positive in SOCA and negative in STC).
Here we show that this disparity originates from the different powers of $k$ in the viscous damping term ($\propto k^2$ in SOCA) and the eddy-damping term ($\propto k^\lambda$ in STC), and can change sign at large Reynolds numbers.
Conversely, the magnetic contribution is much less sensitive to the Reynolds numbers.

To reveal the roles played by the viscous and the eddy-damping terms, it is sufficient to work perturbatively with small shear rate, in which case
\beq
\mathcal{G}_{ijab}=\rho\delta_{ia}\delta_{jb}
-\rho^2\mathcal{S}_{ijab}+\mathcal{O}(S^2),
\label{eqn:G_small_sh}
\eeq
where
\beq
\rho(k)=\frac{1}{\nu_1 k^2+k^{\lambda}},
\eeq
and
\beq
\mathcal{S}_{ijab}=S \delta_{j2}\delta_{ia}\delta_{1b}
-S (2P_{i2}-\delta_{i2})\delta_{1a}\delta_{jb}.
\label{eqn:S4}
\eeq
One can verify that $\mathcal{N}_{ijab}\mathcal{G}_{abmn}=\delta_{im}\delta_{in}+\mathcal{O}(S^2)$.
For simplicity, we have here used $\tau=\kf=1$, $\nu_1=\nu+\nuM$, and dropped the $\Sh$ dependence of the eddy-damping term, which will not affect our analysis.
In these units we have $\nu_1=\ReN^{-1}+\ReM^{-1}$.

Using Equation (\ref{eqn:G_small_sh}) in Equation (\ref{eqn:Cab}) and keeping terms up to oder $\mathcal{O}(S)$ we obtain
\begin{align}
\tilde C_{ab}=&\left[
\tilde T^\Lambda_{\alpha\beta}
+\frac{1}{2}\Lambda_{nm}k_m\left(\tilde K_{\alpha\beta}+\tilde M_{\alpha\beta}\right)\frac{\partial}{\partial k_n}\right]\notag\\
&\times
\left[
\left(\rho+\frac{Sk_1k_2}{k^2}k\rho\rho'\right)\delta_{a\alpha}\delta_{b\beta}
-\rho^2\mathcal{S}_{ab\alpha\beta}
\right]\notag\\
&+\text{total derivatives}+\mathcal{O}(S^2),
\label{eqn:perturbCij}
\end{align}
where $\rho'\equiv \partial \rho/\partial k$,
and the $\beta_{21}$ coeffcient can be shown to be
\begin{align}
&\beta_{21}=\frac{8\pi\Sh}{15}
\int k^2dk\ \left[
I^uE^u+I^bE^b
\right],
\label{eqn:b21_perturb}\\
&I^u=-k\rho\rho'-\rho^2,\\
&I^b=-3\rho^2.
\end{align}
If we had ignored the pressure gradient term rather than incorporating it by use of the incompressible condition and the projection operator, we would have replaced $k_ik_j/k^2\to0$ everywhere in Equation (\ref{eqn:perturbCij}) except those in $\tilde K_{ij}$ and $\tilde M_{ij}$, for all $i,j=1,2,3$, and obtained $\beta_{21}=0$.
This is in agreement with \cite{Squire2016jpp}.

We now determine the signs of the kinetic and magnetic contributions to Equation (\ref{eqn:b21_perturb}).
First, 
\beq
\int k^2dk\ I^uE^u=
\int dk\ k^4\rho^3 E^u\left[k^{\lambda-2}(\lambda-1)+\nu_1\right],
\label{eqn:IuEu}
\eeq
so that
\beq
\sgn{I^u}=\left\{\begin{aligned}
&1 & \nu_1\to\infty \\
&\sgn{\lambda-1} & \nu_1\to0
\end{aligned}\right\}.
\label{eqn:sgn_Iu}
\eeq
Next, regardless of the Reynolds numbers,
\beq
\sgn{I^b}=-1.
\label{eqn:sgn_Ib}
\eeq
The kinetic contribution is positive at small Reynolds numbers, but it can change its sign if the correlation time for $\tilde C_{ij}$ scales 
with the wavenumber not too steeply, namely $\lambda<1$ or equivalently $\qspec<2$. 
On the other hand, the sign of the magnetic contribution is more robust, being consistently negative regardless of the Reynolds numbers and spectral index.
These conclusions are consistent with Equation (25) of \cite{RogachevskiiKleeorin2004} which applies for large Reynolds numbers.
This different behavior of the kinetic and magnetic SCE arises because the projection operator $\hat P_{ij}$ is applied to the Navier-Stokes equation but not the small-scale induction equation, 
consistent with \cite{Squire2016jpp}.

\begin{figure*}
\centering
\includegraphics[width=2\columnwidth]{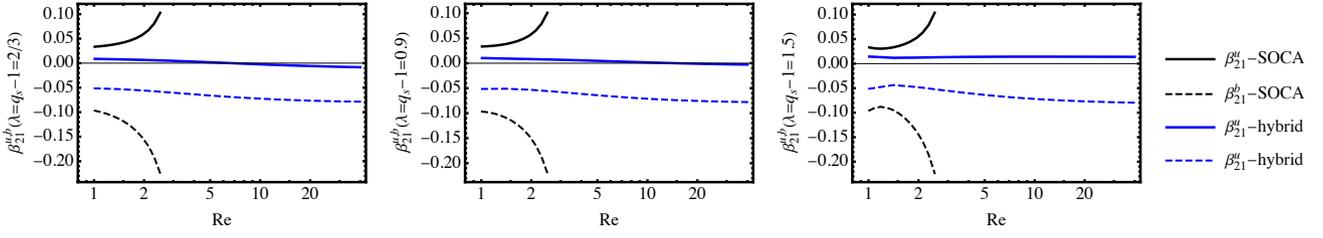}
\caption{%
The kinetic and the magnetic contributions to $\beta_{21}$ from SOCA and STC with dissipation (``hybrid'') closures.
$\beta^u_{21}$ remains positive in SOCA, but reverses its sign as $\ReN$ increases in the presence of an eddy-damping term in STC.
$q_\text{spec}$ labels the spectral index;
$q_\text{spec}=5/3$ corresponds to a Kolmogorov-type spectrum.
}
\label{fig:nu-P1R0S1}
\end{figure*}

These properties manifest when we solve for $\beta_{21}$ numerically at order $\mathcal{O}(\Sh^3)$, as shown in Figure \ref{fig:nu-P1R0S1}.
The values of $\beta^{u,b}_{21}$ are computed at $\PrM=1$ and $\Sh=0.3$, with different choices of $\lambda=\qspec-1$.
For a Kolmogorov-type spectrum ($\qspec=5/3$ or $\lambda=2/3$), the sign transition of $\beta^u_{21}$ happens at $\ReN\simeq6$, and this critical Reynolds number becomes $\sim15$ for a steeper spectrum $\qspec=1.9$.
For $\qspec=2.5$, the kinetic contribution remains positive for all $\ReN$.
As for the magnetic contribution, SOCA and STC approaches both give a robust negative sign for all Reynolds numbers.

\begin{figure*}
\centering
\includegraphics[width=2\columnwidth]{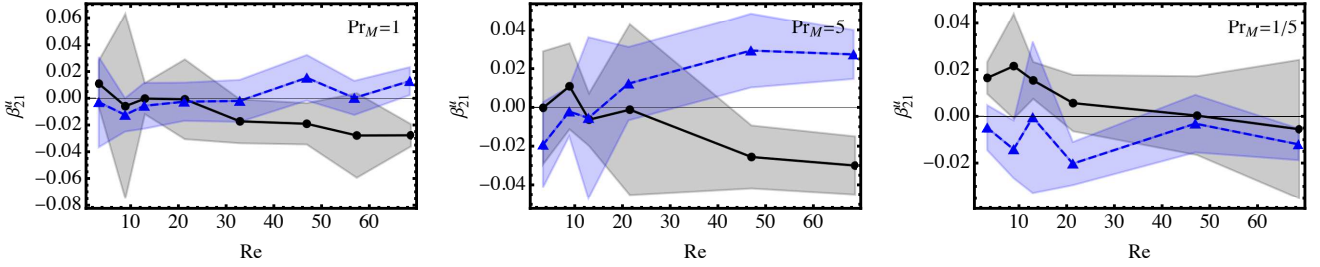}
\caption{%
Measuring the kinetic SCE $\beta^u_{21}$ in kinematic TFM.
For all data points, $\Sh\simeq0.07$.
Solid black curves: shear flow without rotation.
Dashed blue curves: shear flow with a Keplerian rotation, so the Coriolis number is $\Co\simeq0.1$ for all data points. 
Shaded areas represent uncertainties calculated from an ensemble of $\sim15$ members, each being an average of $6$ measurements.}
\label{fig:KTFM}
\end{figure*}

\subsection{The rotating case}
To include the Coriolis force, we add an extra contribution from rotation to Equation (\ref{eqn:S4}),
\beq
\Delta\mathcal{S}_{ijab}=\frac{2S}{q}P_{im}\epsilon_{m3a}\delta_{jb},
\eeq
and we then have
\beq
I^u=\frac{2-q}{q}k\rho\rho'+\frac{3-2q}{2q}\rho^2,
\eeq
and
\beq
I^b=\frac{2}{q}k\rho\rho'+\frac{3(3-2q)}{2q}\rho^2.
\eeq
The non-rotating limit can be recovered by taking $q=S/\Omega\to\infty$.
For the special case of a Keplerian rotation $q=3/2$, we have $I^u=I^b/4=k\rho\rho'$,
and therefore
\beq
\sgn{I^{u,b}}=\left\{\begin{aligned}
&-1 & \nu_1\to\infty \\
&-\sgn{\lambda} & \nu_1\to0
\end{aligned}\right\}.
\label{eqn:sgn_Iub_rot}
\eeq
Comparing Equation (\ref{eqn:sgn_Iub_rot}) with Equations (\ref{eqn:sgn_Iu}) and (\ref{eqn:sgn_Ib}) we recover the conclusion of \cite{Squire2015pre} that adding Keplerian rotation will turn $\beta^u_{21}$ from positive to negative values in SOCA.
Furthermore, the limiting case when $\nu_1\to0$ also implies that although a shallow spectrum with $\lambda=\qspec-1<0$ facilitate the kinetic SCE in the non-rotating case, it does the opposite when Keplerian rotation is present.

Table \ref{tab:beta21} summarizes the key results of this Section.

\begin{table}
\caption{Signs of kinetic and magnetic $\beta_{21}$ coefficients from SOCA and STC for non-rotating and rotating turbulence.
Here $\lambda=-\ln\taucor/\ln k$ captures how the turbulent correlation time scales with the wavenumber.
A typical choice is $\lambda=\qspec-1$ where $\qspec$ is the spectral index of the kinetic energy.
}
\label{tab:beta21}
\begin{tabular}{cccc}
\hline
Turbulence with shear & Closure & $\beta^u_{21}$ & $\beta^b_{21}$ \\
\hline
\multirow{2}{*}{No rotation} & SOCA & $>0$ & $<0$ \\
 & STC & $\sgn{\lambda-1}$ & $<0$ \\
\hline
\multirow{2}{*}{Keplerian rotation} & SOCA & $<0$ & $<0$ \\
 & STC & $-\sgn{\lambda}$ & $-\sgn{\lambda}$ \\
\hline
\end{tabular}
\end{table}

\section{Implication for finding SCE in simulations}
\label{sec:simulation}
\subsection{The kinetic contribution}
At low Reynolds numbers, \cite{Squire2015apj} ($\ReN=\ReM\simeq5$) found simulations to agree with the SOCA theory \citep{Squire2015pre} where $\beta^u_{21}$ is positive with only shear, and negative when a Keplerian rotation is added.
\cite{Brandenburg2008magdiff} ($\ReN=1.4,\ReM=14$), \cite{SinghJingade2015} ($\text{min}\left\{\ReN,\ReM\right\}<1$), and \cite{Kapyla2020} ($\ReN\simeq0.5, \ReM\lesssim10$) all supported the former case, but did not explore the rotation case.
Notably, some of these simulations have a Reynolds exceeding the critical value ($\sim15$) we found in the previous section, yet still found a positive $\beta^u_{21}$.
This is possibly due to the steeper energy spectra at low Reynolds numbers which, if spectral index
 $\qspec>2$, generates a positive $\beta^u_{21}$.
At higher Reynolds numbers ($\ReN\gtrsim100$), there has not been a thorough numerical investigation, even for the kinetic SCE, to compare with results from STC.
Below we provide preliminary evidence of the kinetic SCE in such turbulent regimes.

We use the kinematic TFM implemented in the \texttt{Pencil Code} \citep{pc2021} to measure exclusively the kinetic SCE coefficient $\beta^u_{21}$;
for details of the kinematic method see \cite{Brandenburg2008magdiff}.
For the magnetic counterpart the fully nonlinear TFM has only been recently established \citep{Kapyla2020,Kapyla2021CTFM}, and may become a useful tool for future investigation.

We first study cases with a shear flow but no rotation.
For all runs, the shear parameters are $\Sh\simeq0.07$.
The solid black curves in Figure \ref{fig:KTFM} show the evidence for $\beta^u_{21}$ transitioning from positive to negative with increasing $\ReN$ for magnetic Prandtl numbers $\PrM=1$, $5$, and $1/5$.
The exact transition point is difficult to pin down because of the sizeable error bars, but there is an indication of larger critical $\ReN$ at smaller $\PrM$, consistent with our theory.
However, for all the pure shear runs presented, the growth rate computed from the transport coefficients from the TFM according to a shear-current dynamo model is negative, meaning that it is present but the SCE is too weak to drive a large-scale dynamo at $\Sh=0.07$ and scale separation $1/5$.

For shearing and rotating cases, we use a Keplerian rotation so that $\Sh\simeq0.07$ and $\Co\simeq0.1$ for all runs.
These are shown in blue dashed curves in Figure \ref{fig:KTFM}.
In general, at low Reynolds numbers the rotation lowers $\beta^u_{21}$, in agreement with \cite{Squire2015apj}. At large $\ReN$ it does the opposite, turning $\beta^u_{21}$ from negative to positive values compared to the pure shear cases.
Both results are in qualitative agreement with our theoretical analysis in the previous Section, although some quantitative inconsistencies are discussed below.

Indeed, at the Reynolds numbers that we have investigated, the spectral index at $k>\kf$ is always larger than $2$, or equivalently, $\lambda>1$.
The theoretical predictions in Table \ref{tab:beta21} then seemingly suggest that for the runs with large Reynolds numbers, $\beta^u_{21}$ would neither be negative for pure shear runs, nor would it be positive for Keplerian runs;
yet, there is evidence for both in simulations.
In fact, naively applying Table \ref{tab:beta21} to Figure \ref{fig:KTFM} would suggest $\lambda<0$ for rotating shearing cases, which is physically unsound given that $\taucor(k) \propto k^{-\lambda}$.

A plausible explanation of this discrepancy is that 
the positive spectral slope at $1<k<\kf$ yields an effective $\lambda$ less than its value when dominated by $k>\kf$ modes;
physically, the effective $\lambda$ being negative implies that the turbulent transport coefficients measured are really dominated by large-scale motions.
$\beta^u_{21}$ then becomes negative for pure shear flows or positive for Keplerian flows, even if $\qspec$ slightly exceeds $2$.
This effect was eliminated from our theoretical calculations, thereby isolating one specific process that can be present in the simulations that is absent in the theory. 
The effect may also produce a larger magnitude of $\beta^u_{21}$ than that in the theory (c.f. Figure \ref{fig:KTFM} and the bottom left panels of Figures \ref{fig:P1R0S1} and \ref{fig:PR}).

The spectral slope near the forcing scale $\kf$ also influences $\beta^u_{21}$. With single-scale forcing the resulting spectrum is not smooth, and usually peaks at the forcing scale.
The spectrum is thus rather steep at $k\gtrsim\kf$, and only after cascading for one or two wavenumbers does its slope approaches a constant.
The integral (\ref{eqn:b21_perturb}), however, will pick up the slope near the energy-dominant wavenumber, which can differ drastically from that in the inertial range.
This effect persists even for simulations with very large Reynolds numbers, but it is likely less influential the more extended the inertial range. 

Finally, we note that the $\qspec$-dependence of $\beta^u_{21}$ essentially originates from its association with the spatial inhomogeneity of $\bmB$, but the converse statement is not true:
Not every transport coefficient associated with $\del\bmB$ depends sensitively on $\qspec$.
The R\"adler ($\bm\Omega\times\bmJ$) effect is a counter-example as it is
associated with the anti-symmetric part of $\Lambda_{ij}$, for which the $k$-derivative of the energy spectra in $\tilde T^B_{ij}$ does not contribute.
To see this, notice that
\begin{align}
\tilde T^B_{ij}=&-\frac{1}{2}\Lambda_{nm}k_m
\frac{\partial}{\partial k_n}\left(
\tilde K_{ij}+\tilde M_{ij}\right)\notag\\
=&-\frac{1}{2}\Lambda_{nm}k_m
\left[\frac{k_n}{k}\frac{\partial R}{\partial k}A_{ij}
+R\frac{\partial}{\partial k_n} A_{ij}
\right]
\end{align}
under the assumption that $(\tilde K_{ij}+\tilde M_{ij})$ can be written as the product of a radial part $R(k)$ and an angular part $A_{ij}(\bmk/k)$.
The sign of the term proportional to $\partial R/\partial k$ will depend on $\qspec$,
but it does not contribute to the part of $\tilde T^B_{ij}$ that is anti-symmetric in $\Lambda_{nm}$.
As such, the R\"adler effect is relatively insensitive to $\qspec$ 
 and can be detected in a broad range of simulations \citep[e.g.,][]{Brandenburg2008magdiff}.

\subsection{The magnetic contribution}
\label{sec:bc}
The numerical results on the magnetic contribution $\beta^b_{21}$ are more disparate and discrepant with theory, although not necessarily mutually contradictory, as the numerical experiments have not been mutually standardized.
\cite{Squire2015apj} ($\ReN=\ReM\simeq5$) used quasi-linear methods in magnetically forced simulations and found agreement with the SOCA theory that $\beta_{21}<0$, either with or without Keplerian rotation.
\cite{Kapyla2020} studied non-linear TFM in MHD burgulence (i.e. ignoring the thermal pressure gradient) with both kinetic and magnetic forcing, and found $\beta_{21}>0$. Although this supports the conclusion of \cite{Squire2016jpp} that the pressure gradient is necessary for magnetic SCE, it actually contradicts theory which predicts $\beta_{21}=0$.

On the other hand, projection methods generally support the theory that $\beta^b_{21}<0$ \citep{Squire2015prl,Shi2016}, albeit imposing additional and probably artificial constraints on $\beta_{xy}$, which threatens self-consistency.

The absence of the SCE in simulations with low hydro and magnetic Reynolds numbers disagrees with theory, especially for those with strong magnetic fluctuations.
In particular, Figure \ref{fig:nu-P1R0S1} shows that, for both SOCA and STC, the magnetic contribution to SCE is larger than the kinetic SCE at energy equipartition $\abra{b^2}=\abra{u^2}$.
Thus in magnetically forced simulations where $\abra{b^2}>\abra{u^2}$, one might have expected $\beta_{21}<0$ as it is dominated by the magnetic contribution.
We speculate that this discrepancy between theory and simulations arises for two reasons:
\begin{enumerate}
\item
When $\qspec>2$, Equation (\ref{eqn:IuEu}) implies that $\beta^u_{21}$ remains positive but is larger in magnitude for a steeper kinetic energy spectrum.
Strictly speaking, when $\qspec\geq3$ the STC formalism is invalid, but the fact that $\beta^u_{21}$ is more positive for steeper spectra seems to still hold.

The left panel of Figure \ref{fig:b21_qspec} illustrates this. There we plot both kinetic and magnetic contributions versus spectral index $\qspec$ in the STC closure with microscopic dissipation.
We also fix $\knu=2\kf$ (otherwise $\qspec\geq3$ cases cannot be self-consistent).
At a critical value $\qspec\simeq4.5$, corresponding to a steep but not uncommon spectrum in low-$\ReN$ simulations, the magnitude of the kinetic contribution becomes comparable to the magnetic contribution.
In the right panel we further explore how this critical spectral index depends on the energy ratio $\epsilon=\abra{b^2}/\abra{u^2}$ in our model. The plotted solid curve corresponds to vanishing $\beta_{21}$, or equivalently $\beta_{21}^u+\epsilon\beta_{21}^b=0$. Even at $\epsilon=2$, $\beta_{21}>0$ if $\qspec\gtrsim5.5$, which is not impossible at $\ReN\lesssim10$.
In other words, very steep kinetic energy spectra in low $\ReN$ simulations may result in a much larger $\beta^u_{21}$ that dominates $\beta_{21}$ even at energy super-equipartition $\epsilon\gtrsim1$.
However, Figure \ref{fig:b21_qspec} should not be interpreted as an exact prediction because of the simplifications, and using STC at such low Reynolds numbers.

\begin{figure*}
\centering
\includegraphics[width=2\columnwidth]{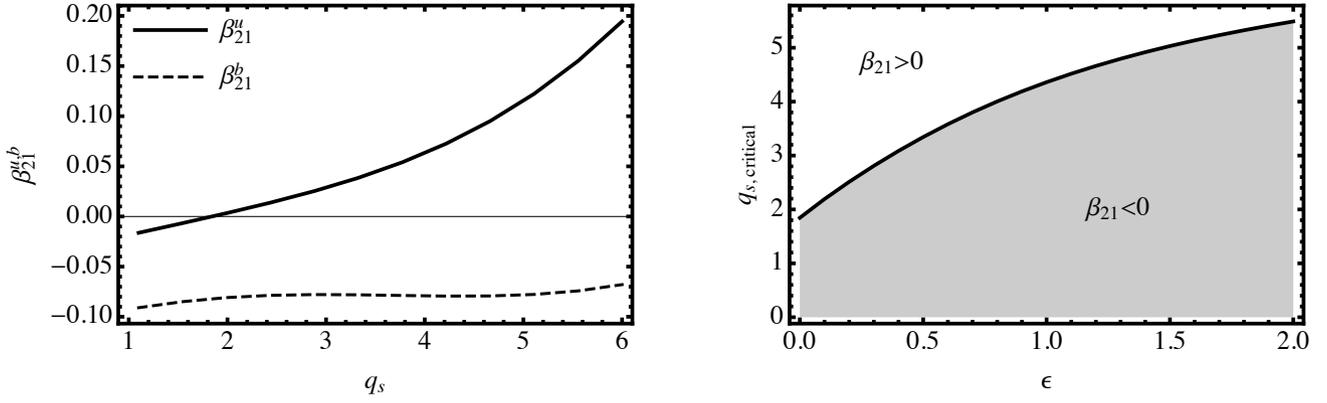}
\caption{%
Left:
The kinetic and the magnetic contributions to $\beta_{21}$ from STC with dissipation (``hybrid'') closure, at $\ReN=\ReM=10$ and $\Sh=0.3$.
$\qspec$ labels the spectral index;
$\qspec=5/3$ corresponds to a Kolmogorov-type spectrum.
Right:
The critical spectral index $q_\text{s,critical}$ at which $\beta_{21}=0$, as a function of energy ratio $\epsilon=\abra{b^2}/\abra{u^2}$.
}
\label{fig:b21_qspec}
\end{figure*}

\item
Another glaring but seldom discussed incongruence between theories and simulations lies in their mutually distinct boundary conditions.
The shearing-box approximation has become a standard set-up for simulating a local differentially rotating patch of system, and the shear-periodic boundary condition is commonly adopted.
On the other hand, most theoretical calculations have used normally periodic conditions, which allows for a straightforward Fourier transform, simplifying the calculations.
An exception is the work using shearing coordinates or a shearing-wave basis \citep{SridharSubramanian2009,SridharSingh2010,SinghSridhar2011}, which agrees with SOCA for the kinetic SCE, but there is no equivalent study for the magnetic counterpart.
Would the shear-periodic or normally periodic boundary condition make a difference for $\beta^b_{21}$?
We speculate that it may, because the magnetic contribution to $\beta^b_{21}$ has been shown in Section \ref{sec:Re_dependence} and in \cite{Squire2016jpp} to result entirely from the pressure gradient term in the incompressible Navier-Stokes equation, at least when normal periodic boundary conditions are used.
More precisely, the magnetic contribution originates from the projection operator $\hat P_{ij}=\delta_{ij}-\partial_i\partial_j/\partial^2$ used to eliminate the pressure term (in both SOCA and STC calculations), where the inverse of the Laplacian operator $\partial^{-2}$ is involved.
Since the results depend on solving Laplace's equation, the extent to which using shear-periodic versus periodic boundary conditions changes the solution and the sign of $\beta^b_{21}$, is unclear. Using the shearing-frame technique in a shearing box recovers normal periodic boundary conditions, but still leaves unresolved how the equations can be solved at large Reynolds numbers.
\end{enumerate}

\section{The full diffusivity tensor and dynamo growth rates}
\label{sec:result}
In what follows we will present the full solution of Equation (\ref{eqn:emf_eta}) by performing the integrals numerically at $\ReN=\ReM=1000$.
We also use $\qspec=5/3$ and $q=3/2$ everywhere.
We are interested in whether the kinetic and the magnetic contributions favor the SCE in three cases:
(i) only shear, no rotation,
(ii) only shear, no rotation, and no total pressure gradient (thus no projection operator acting on the Navier-Stokes equation),
(iii) both shear and rotation.
In each case, we present the coefficients expanded in Taylor series with small $\Sh$ (or the Coriolis number $\Co$) as well as the full solution plot for $0\leq\Sh\leq0.5$.
The coefficients should be interpreted as being in the kinematic phase when the mean magnetic field is weak.
As pointed out in Appendix \ref{appx:corfunc}, the assumption of isotropic background turbulence only holds up to $\Ro\simeq2$, corresponding to $\Co\lesssim0.5$ or $\Sh\lesssim 0.375$ with a Keplerian rotation.
When the rotation is present, we additionally split $\Lambda_{ij}$ into symmetric and anti-symmetric parts, so that the R\"adler effect can be isolated by considering only the anti-symmetric part of $\Lambda_{ij}$.

\subsection{Shear only, no rotation}
For the case of a shear flow and pressure gradient in the Navier-Stokes equation, but without rotation, we find at small $\Sh$\begin{align}
\beta_{ij}\simeq&
\frac{\tau\abra{u^2}}{3}
\begin{pmatrix}
1-0.8\Sh & 0.64\Sh \\
-0.04\Sh & 1-0.8\Sh
\end{pmatrix}\notag\\
&+\frac{\tau\abra{b^2}}{3}\begin{pmatrix}
0 & 0.02\Sh \\
-0.4\Sh & 0
\end{pmatrix}
+\mathcal{O}(\Sh^2).
\end{align}
This agrees with \cite{RogachevskiiKleeorin2004} albeit our values of $\beta^u_{21}$ and $\beta^b_{21}$ are $\sim4/3$ times larger than theirs.

\begin{figure*}
\centering
\includegraphics[width=2\columnwidth]{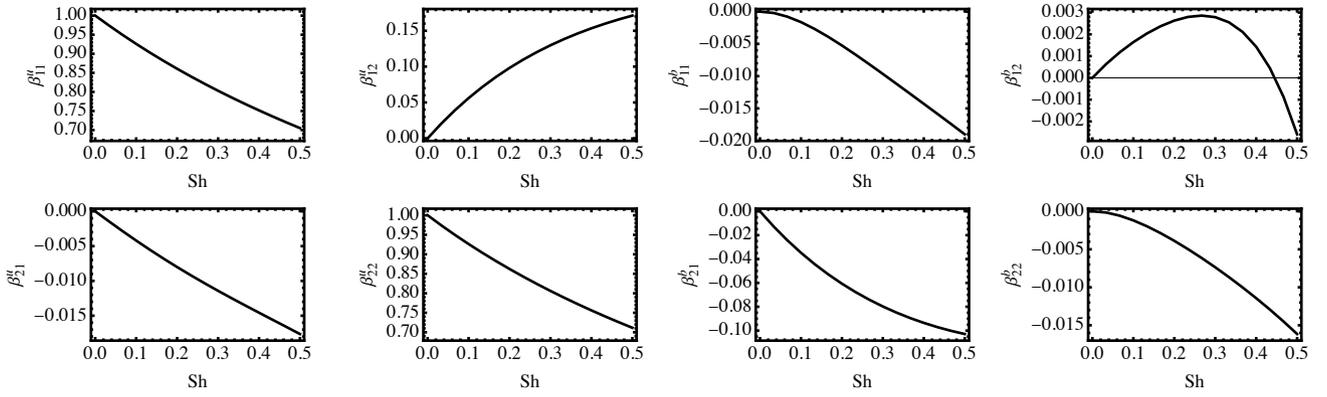}
\caption{Diffusion coefficients in a homogeneous turbulence with shear and pressure gradient, but no rotation.}
\label{fig:P1R0S1}
\end{figure*}

The full solution of $\beta_{ij}$ is shown in Figure \ref{fig:P1R0S1}.
In general, we confirm both kinetic and magnetic SCE in our calculations, in agreement with previous STC calculations \citep{RogachevskiiKleeorin2003,RogachevskiiKleeorin2004,Pipin2008}.
When comparing with Figures 3 and 4 of \cite{Brandenburg2008magdiff}, we find general agreement that
$\beta_{12}^u$ increases with increasing $\Sh$, and saturates to non-zero values at large $\ReM$.
We also notice that $|\beta^u_{12}|$ is comparable to $|\beta^b_{21}|$, which would seem to invalidate the method of \cite{Squire2015prl} and \cite{Shi2016} who set $\beta_{21}=0$ in the projection method.
In the previous section we discussed the contradictory signs of $\beta^u_{21}$ found with STC and SOCA respectively, and speculated as to why simulations found $\beta_{21}>0$ whereas theory predicted $\left|\beta^b_{21}\right|>\left|\beta^u_{21}\right|$ at energy equipartition.

\subsection{Shear only, no rotation, and no pressure}
\begin{figure*}
\centering
\includegraphics[width=2\columnwidth]{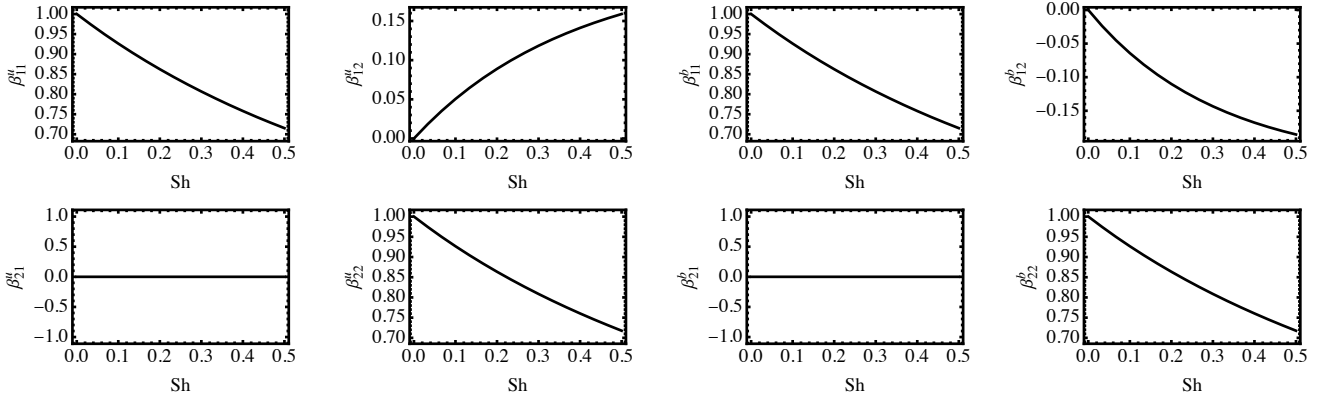}
\caption{Similar to Figure \ref{fig:P1R0S1}, but in a homogeneous turbulence with shear, without rotation, and without the pressure gradient term in the Navier-Stokes equation.}
\label{fig:nP1R0S1}
\end{figure*}

For turbulence with shear but without rotation and total pressure gradient, we find
\begin{align}
\beta_{ij}\simeq&
\frac{\tau\abra{u^2}}{3}
\begin{pmatrix}
1-0.8\Sh & 0.58\Sh \\
0 & 1-0.8\Sh
\end{pmatrix}\notag\\
&+\frac{\tau\abra{b^2}}{3}\begin{pmatrix}
1-0.8\Sh & -0.76\Sh \\
0 & 1-0.8\Sh
\end{pmatrix}
+\mathcal{O}(\Sh^2).
\end{align}
\cite{Squire2016jpp} argued that the pressure gradient term, $\del p$, in the Navier-Stokes equation is necessary for the SCE in the SOCA approach.
We confirm that this is also the case at large Reynolds numbers in STC, as presented in Figure \ref{fig:nP1R0S1}:
both $\beta^u_{21}$ and $\beta^b_{21}$ vanish once we removed $\del p$.
However, simulations have not yet provided a direct comparison with the regime of our theoretical results in this regard. 
\cite{Kapyla2020} studied a MHD burgulence (simplified MHD), but used small hydro and magnetic Reynolds numbers.
The presence of finite $\beta^b_{11,22}$ without a pressure gradient has been well-known.

\subsection{Shear with Keplerian rotation}
For homogeneous turbulence with both shear and Keplerian rotation, we find
\begin{align}
\beta_{ij}\simeq&
\frac{\tau\abra{u^2}}{3}
\begin{pmatrix}
1-0.7\Sh & 0.45\Sh \\
-0.33\Sh & 1-0.7\Sh
\end{pmatrix}\notag\\
&+\frac{\tau\abra{b^2}}{3}\begin{pmatrix}
-0.1\Sh & -0.11\Sh \\
0.006\Sh & -0.1\Sh
\end{pmatrix}\notag\\
&+
\frac{\tau\abra{u^2}}{3}
\begin{pmatrix}
-0.1\Sh & 0.18\Sh \\
0.30\Sh & -0.1\Sh
\end{pmatrix}\notag\\
&+\frac{\tau\abra{b^2}}{3}\begin{pmatrix}
0.1\Sh & -0.15\Sh \\
-0.13\Sh & 0.1\Sh
\end{pmatrix}
+\mathcal{O}(\Sh^2).
\end{align}
Here, the first and second lines are the contributions from the anti-symmetric part of $\Lambda_{ij}$, and the third and forth lines are the symmetric part.
Thus, the $\beta^u_{12}>0$ and $\beta^u_{21}<0$ values in the first line manifest the R\"adler effect.

\begin{figure*}
\centering
\includegraphics[width=2\columnwidth]{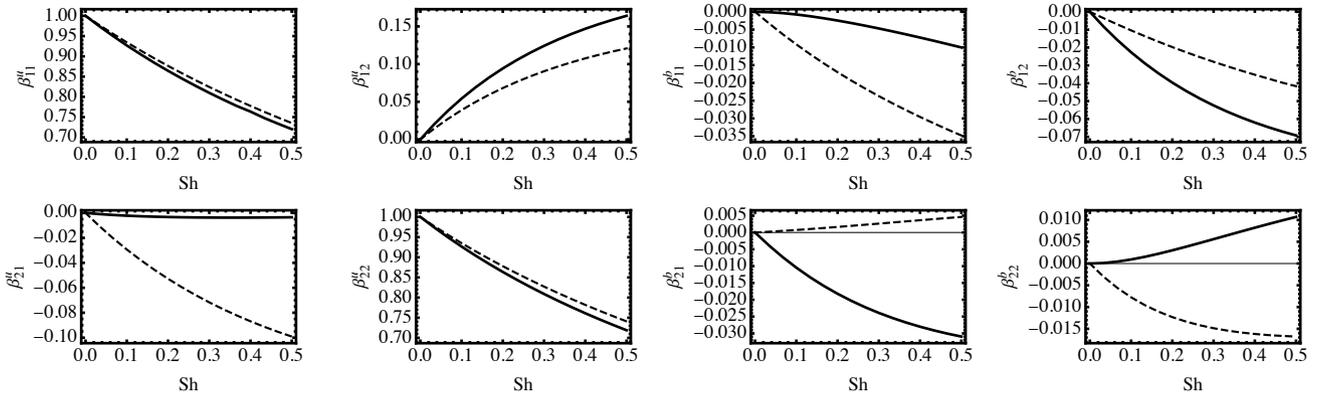}
\caption{Similar to Figure \ref{fig:P1R0S1}, but in a homogeneous turbulence with both shear and a Keplerian rotation.
The Rossby number is given by $\Ro=3/(4\Sh)$.
The dashed curves show the part that gives the R\"adler effect.
}
\label{fig:PR}
\end{figure*}

In Figure \ref{fig:PR} we show $\beta_{ij}$ for $\Sh\leq0.5$, and the Coriolis number is $\Co=2\Sh/q=4\Sh/3$ for all panels.
The dashed curves show the contribution from the anti-symmetric part of $\Lambda_{ij}$, and the solid curves show the full solution.
In this STC regime, adding a Keplerian rotation suppresses both kinetic and magnetic shear-current effects (c.f. Figure \ref{fig:P1R0S1}), opposite to the SOCA regime where $\beta^u_{21}$ changes from positive to negative when the rotation is added \citep{Squire2015apj,Squire2015pre}.

\subsection{Rotating turbulence without shear}
For completeness, we present the results when there is no shear but only rotation:
\begin{align}
\beta_{ij}
\simeq&
\frac{\tau\abra{u^2}}{3}
\begin{pmatrix}
1-0.52\Co & 0.16\Co \\
-0.16\Co & 1-0.52\Co
\end{pmatrix}\notag\\
&+\frac{\tau\abra{b^2}}{3}\begin{pmatrix}
-0.07\Co & -0.17\Sh \\
0.17\Sh & -0.07\Co
\end{pmatrix}\notag\\
&+
\frac{\tau\abra{u^2}}{3}
\begin{pmatrix}
-0.07\Co & -0.17\Co \\
0.17\Co & -0.07\Co
\end{pmatrix}\notag\\
&+\frac{\tau\abra{b^2}}{3}\begin{pmatrix}
0.07\Co & -0.04\Co \\
0.04\Co & 0.07\Co
\end{pmatrix}
+\mathcal{O}(\Co^2),
\end{align}
or, for the convenience of comparison, in term of a formal shear parameter $\Sh=3\Co/4$ (note that there is no shear),
\begin{align}
\beta_{ij}\simeq&
\frac{\tau\abra{u^2}}{3}
\begin{pmatrix}
1-0.69\Sh & 0.22\Sh \\
-0.22\Sh & 1-0.69\Sh
\end{pmatrix}\notag\\
&+\frac{\tau\abra{b^2}}{3}\begin{pmatrix}
-0.10\Sh & -0.23\Sh \\
0.23\Sh & -0.10\Sh
\end{pmatrix}\notag\\
&+
\frac{\tau\abra{u^2}}{3}
\begin{pmatrix}
-0.10\Sh & -0.23\Sh \\
0.23\Sh & -0.10\Sh
\end{pmatrix}\notag\\
&+\frac{\tau\abra{b^2}}{3}\begin{pmatrix}
0.10\Sh & -0.05\Sh \\
0.05\Sh & 0.10\Sh
\end{pmatrix}
+\mathcal{O}(\Sh^2).
\end{align}
Again, in each of the two previous equations, 
 the first and second lines come from the anti-symmetric part of $\Lambda_{ij}$, and the third and forth lines the symmetric part.

\begin{figure*}
\centering
\includegraphics[width=2\columnwidth]{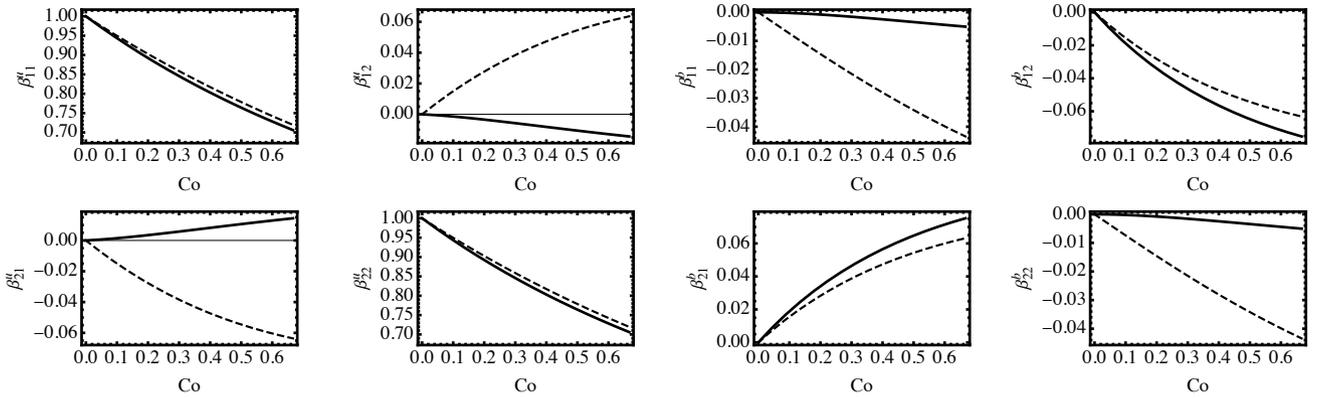}
\caption{Similar to Figure \ref{fig:P1R0S1}, but in a homogeneous turbulence with only rotation, no shear.
The dashed curves represent the R\"adler effect.
}
\label{fig:PoR}
\end{figure*}

The numerical solution is shown in Figure \ref{fig:PoR}, with dashed curves showing the contribution from the anti-symmetric part of $\Lambda_{ij}$ and solid curves showing the full solution.
The R\"adler effect makes the amplitudes of the negative kinetic off-diagonals much smaller, and can even change their signs, whereas the magnetic off-diagonals remain roughly unchanged \citep{Raedler2003,BrandenburgSubramanian2005,ChamandySingh2017}.
The results resemble the ``magnetic R\"adler effect'' for planar-averaged mean fields, which is the combination of the R\"adler effect and the $\kappa$ term [c.f. Equation (\ref{eqn:emf_decomposition})].
Note that there is a disagreement between \cite{Raedler2003} and \cite{BrandenburgSubramanian2005} regarding the value of $\kappa_{ijk}$, as noted by \cite{ChamandySingh2017}.
Here, our results agree more so with those of \cite{BrandenburgSubramanian2005}, with a slight difference in the specific values.

\subsection{Dynamo growth rates}
\label{sec:gamma}
Finally, we explore the dynamo growth rates due to the combined effects of the turbulent diffusivity and the shear, with or without a Keplerian rotation.
The mean-field induction equations are
\begin{align}
&\partial_tB_x=\beta_{22}\nabla^2 B_x-\beta_{21}\nabla^2 B_y,\\
&\partial_tB_y=-SB_x+\beta_{11}\nabla^2 B_y-\beta_{12}\nabla^2 B_x,
\label{eqn:dtby}
\end{align}
assuming that the turbulent diffusivity $\beta_{11,22}$ are much larger than the microscopic diffusivity $\nuM$.
The dynamo growth rate is
\begin{align}
\gamma=&
\left[\frac{1}{4}(\beta_{11}-\beta_{22})^2K_z^4
+\beta_{12}\beta_{21}K_z^4-S\beta_{21}K_z^2\right]^{1/2}\notag\\
&-\frac{\beta_{11}+\beta_{22}}{2}K_z^2.
\end{align}
where $K_z=|\del\bmB|/|\bmB|$ is the scale of $\bmB$.
Notably, $\gamma$ is either real and positive, or complex but with a negative real part, which implies the absence of an oscillatory growing mode.

\begin{figure*}
\centering
\includegraphics[width=2\columnwidth]{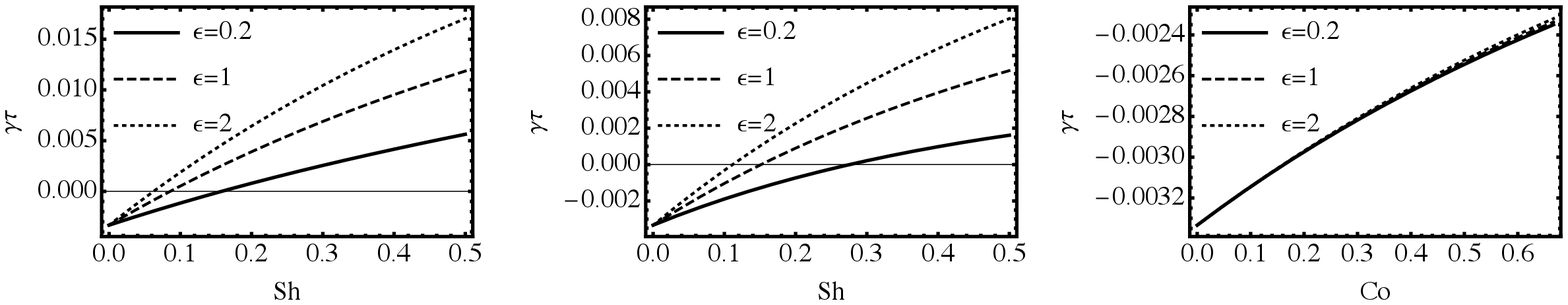}
\caption{
The dynamo growth rates $\gamma$ at scale separation parameter $\zeta=\tau\urms K_z=0.1$ and different levels of the fluctuating magnetic field, $\epsilon=\abra{b^2}/\abra{u^2}$.
Left:
Only shear, no rotation.
Middle:
Both shear and a Keplerian rotation.
Right:
Only rotation, no shear.
}
\label{fig:gamma}
\end{figure*}

Using the energy scaling factor $\epsilon=\abra{b^2}/\abra{u^2}$, the growth rate can be written in dimensionless form,
\begin{align}
\gamma\tau=&
\left[\frac{(\beta_{11}^\epsilon-\beta_{22}^\epsilon)^2}{4}
\frac{\zeta^4}{9}
+\frac{\beta_{12}^\epsilon\beta_{21}^\epsilon\zeta^4}{9}
-\Sh\beta_{21}^\epsilon\frac{\zeta^2}{3}\right]^{1/2}\notag\\
&-\frac{\beta_{11}^\epsilon+\beta_{22}^\epsilon}{2}\frac{\zeta^2}{3},
\end{align}
where $\beta^\epsilon_{ij}=\beta_{ij}^u+\epsilon \beta_{ij}^b$, and $\zeta=\tau\urms K_z$ characterizes the scale separation between the turbulent and the mean fields.
The growth rate $\gamma$ (normalized by $\tau^{-1}$) is plotted in Figure \ref{fig:gamma} for different cases: with only shear (left); with both shear and Keplerian rotation (middle); or with only rotation (right).
A SCE-driven dynamo is manifested in the left panel, and magnetic fluctuations indeed aid the dynamo growth.
However, we did not find any growth mode in the parameter regime reported by \cite{Yousef2008prl}, suggesting the growth in that regime is not dominated by a SCE dynamo.

\begin{figure*}
\centering
\includegraphics[width=1.33\columnwidth]{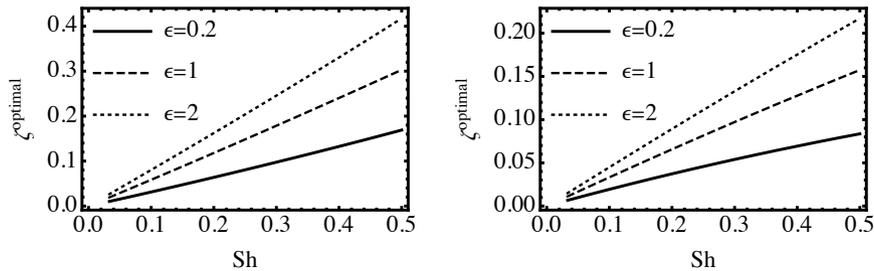}
\caption{
The scale separation parameter $\zeta$ associated with the maximal growth rate.
Left: Only shear, no rotation.
Right: Shear and a Keplerian rotation.}
\label{fig:KzS}
\end{figure*}

\begin{figure*}
\centering
\includegraphics[width=1.33\columnwidth]{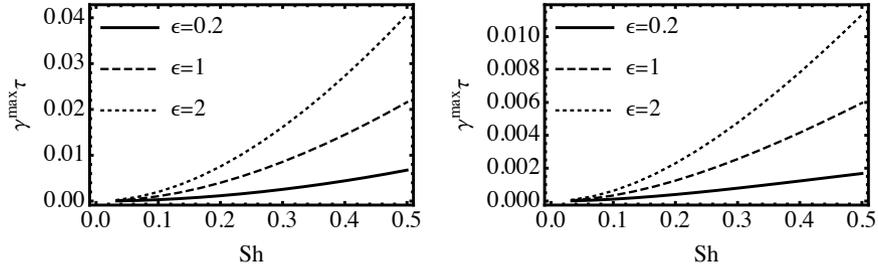}
\caption{
The maximal growth rate if the optimal scale separation is permitted.
Left: Only shear, no rotation.
Right: Shear and a Keplerian rotation.}
\label{fig:gammaOpt}
\end{figure*}

The (dimensionless) optimal mean-field wavenumber associated with the maximal growth rate is shown in Figure \ref{fig:KzS}, exhibiting an approximately linear scaling, $\zeta^\text{optimal}\propto \Sh$.
This agrees with previous STC results \citep{RogachevskiiKleeorin2003,RogachevskiiKleeorin2004} but is against simulations at low Reynolds numbers \citep{Yousef2008prl,Yousef2008AN}, again suggesting a non-SCE origin of the latter dynamos.
The corresponding maximal growth rate $\gamma^\text{max}$ is plotted in Figure \ref{fig:gammaOpt}, showing a faster-than-linear scaling.
Potentially, the $\gamma^\text{max}-\Sh$ relation can be used to distinguish a SCE dynamo from fluctuating $\alpha$ dynamos \citep[$\gamma^\text{max}\propto \Sh$,][]{Jingade2018jpp,JingadeSingh2021}.

\begin{figure*}
\centering
\includegraphics[width=1.33\columnwidth]{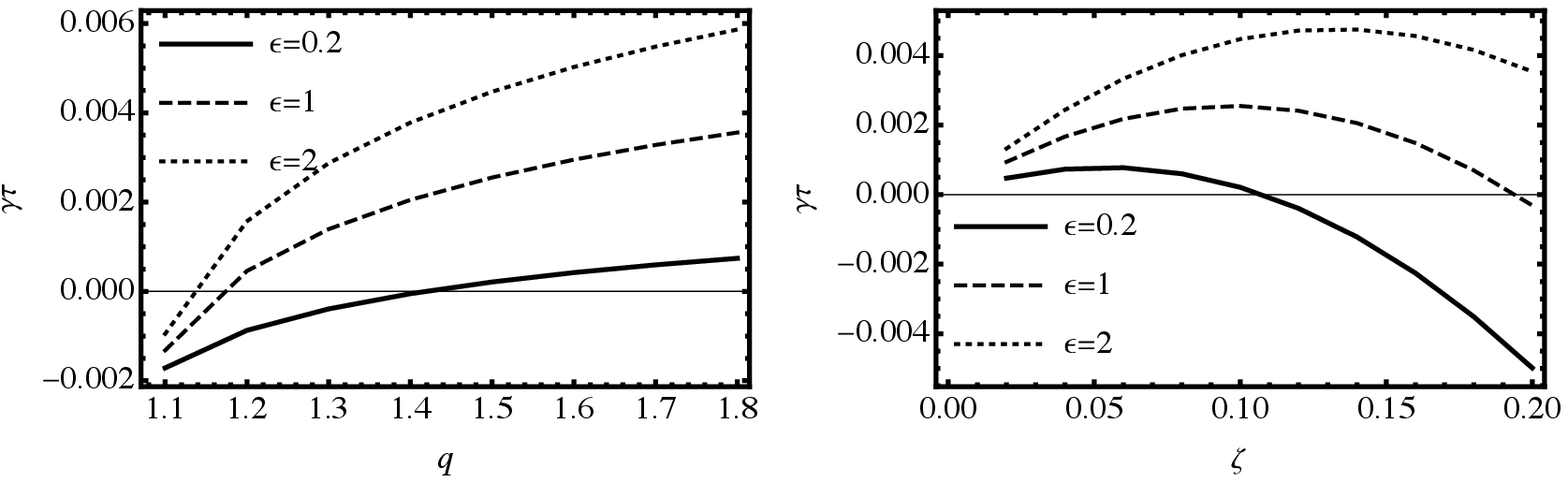}
\caption{
The dynamo growth rates for the case of a rotating turbulence with shear.
For both plots, $\Sh=0.3$.
Left:
Varying the differential rotation parameter $q=-\ln\Omega/\ln r$, at fixed $\zeta=0.1$.
Right:
Varying the scale seperation factor $\zeta$ at fixed $q=3/2$.
}
\label{fig:q}
\end{figure*}

For the case with both shear and rotation, the dependence of the dynamo growth rate on the differential rotation index $q$ and the scale separation parameter $\zeta$ is further explored in Figure \ref{fig:q}.
These relations might not however apply to shearing box simulations driven by the MRI where (i) the turbulence is nonlinearly generated from the shear driven MRI itself and (ii) the turbulence exhibits strong anisotropy.
We nevertheless put them here for reference.

\section{Conclusion}

\label{sec:conclusion}
The first theoretical proposal of the shear-current effect \citep[SCE;][]{RogachevskiiKleeorin2003,RogachevskiiKleeorin2004} used the spectra-$\tau$ closure (STC), and suggested that both the kinetic and the magnetic SCE exist, but later second-order-correlation-approximation (SOCA) \citep{RaedlerStepanov2006,Ruediger2006} and quasi-linear \citep{SridharSubramanian2009,SridharSingh2010,SinghSridhar2011} calculations, as well
as kinematic simulations \citep{Brandenburg2008magdiff}, disagreed with the existence of the kinetic SCE.
Complementarily, the magnetic SCE was confirmed by SOCA \citep{Squire2015pre}, but its existence in simulations is still debated \citep{Squire2015prl,Squire2015apj,Squire2016jpp,Kapyla2020}.

In this work we proposed explanations for both the disagreement between theoretical STC and SOCA results, and between theories and simulations.
Our detailed theoretical investigation reveals that the kinetic SCE coefficient, $\beta^u_{21}$, is always positive if the microscopic diffusion time ($\propto k^{-2}$) is shorter than the correlation time, as is the case in SOCA.
For large Reynolds numbers however, the eddy-damping timescale ($\propto k^{-\lambda}$) is shorter than the viscous time and becomes the correlation timescale. Then, if $\lambda<1$, the kinetic SCE may exist, as the case in STC.
Using the kinematic test-field method (TFM) we demonstrated that $\beta^u_{21}$ transits from positive to negative values as we increase the Reynolds number.
Altogether, we have provided a self-consistent picture to explain the seemingly contradicting results of kinetic SCE among STC, SOCA, and numerical simulations.

The uncertainty of finding a negative $\beta_{21}$ in magnetically forced turbulence simulations \citep{Squire2015apj,Squire2015prl,Squire2016jpp,Kapyla2020} is curious because
 both STC and SOCA theories predict that the magnetic SCE is negative and should dominate the total $\beta_{21}$.
We suggest that the resolution of this apparent contradiction lies again in the distinction of results at different Reynolds numbers 
because $\beta^u_{21}$ depends more sensitively on the kinetic spectral index $\qspec$ than its magnetic counterpart $\beta^b_{21}$. For very steep spectra [$\qspec=\mathcal{O}(5)$] 
typical of low Reynolds numbers $\ReN=\mathcal{O}(10)$, the positive kinetic SCE may then actually dominate even if the small scale magnetic field energy dominates the turbulent kinetic energy. 
We did not explore this numerically because of present limitations of the TFM, but warrants future work.

\cite{Squire2016jpp} 
offered a graphical vector description of how the magnetic fluctuations may give rise to a negative $\beta^b_{21}$ but non-negative $\beta^u_{21}$
 by tracking the relevant terms in the SOCA calculation. 
Whether that picture can be reconciled with the spectral index dependence of the kinetic SCE that we have uncovered here, also warrants investigation.

The role of the thermal pressure gradient term
 raises an interesting 
 question regarding the results in \cite{Kapyla2020}, where MHD burgulence is shown to yield negative $\beta_{21}$ when forced kinetically, and positive $\beta_{21}$ when both kinetic and magnetic energies are forced.
As burgulence lacks the thermal pressure term, both results might seem to contradict the claim of \cite{Squire2016jpp} that $\beta_{21}$ should vanish in this case.
However, since the Lorentz force contributes a magnetic pressure, the MHD burgulence equations will arguably be identical to the MHD turbulence equations if the incompressible condition is used to eliminate the magnetic pressure term.
In this regard, incompressible MHD burgulence should behave in the same manner as incompressible MHD turbulence. Why there is a difference between the two cases seen in \cite{Kapyla2020} (where Mach number is $0.03-0.04$) becomes an interesting question for further work.

Finally, understanding the extent to which what is learned from SCE in forced turbulence can be applied to accretion discs also remains an opportunity for further work. In particular, anisotropy in Reynolds and magnetic stresses is intrinsic in MRI turbulence \citep[e.g.,][]{Pessah2006prl}.
This contrasts typical SCE theories and most simulations that employ a shear flow superimposed on background isotropic turbulence.
There is some evidence for the SCE in shearing-box MRI simulations \citep{Shi2016},
although recently questioned by \cite{Wissing2021},
and whether it influences field growth in global simulations remains to be determined.

\section*{Acknowledgments}
We thank Axel Brandenburg for discussions, and the anonymous referee for useful suggestions.
EGB acknowledges support from grants US Department of Energy DE-SC0001063, DE-SC0020432, DE-SC0020103, and US NSF grants AST-1813298, PHY-2020249.

\section*{Data availability}
The data underlying this article will be shared on reasonable request to the corresponding author.

\bibstyle{mnras}
\bibliographystyle{mnras}

\bibliography{scbib}

\onecolumn

\appendix
\section{Derivation of Equation (29)}
\label{appx:TBij}
Equation (\ref{eqn:TBij}) can be separated into a kinetic and a magnetic contribution,
\begin{align}
&T^{B,u}_{ij}=\dl_m \int d^3x\ 
B_m(\bmx+\bm l)\abra{u_i(\bmx)u_j(\bmx+\bm l)}
,\\
&T^{B,b}_{ij}=-\dl_m \int d^3x\ 
B_m(\bmx)\abra{b_i(\bmx)b_j(\bmx+\bm l)}.
\end{align}
The Fourier transform of its kinetic contribution $T^{B,u}_{ij}$ is
\beq
\tilde T^{B,u}_{ij}
=ik_m\int \frac{d^3k'}{(2\pi)^3}\ 
\tilde B_m(\bmk')
\abra{\tilde u_i^*(\bmk)\tilde u_j(\bmk-\bmk')}.
\eeq
To exploit the scale separation between $\bmB$ and $\bmu$, we notice that the energy spectrum of the mean magnetic field has support primarily from small $|\bmk'|$, which suggests an expansion of $\tilde u_j(\bmk-\bmk')$ at small $\bmk'\ll\bmk$:
\begin{align}
\tilde T^{B,u}_{ij}
&=ik_m\int \frac{d^3k'}{(2\pi)^3}\ 
\tilde B_m(\bmk')\abra{
\tilde u_i^*(\bmk)\tilde u_j(\bmk)
-k'_n\tilde u_i^*(\bmk)\frac{\partial u_j(\bmk)}{\partial k_n}
}\notag\\
&=ik_m\left[
\abra{\tilde u_i^*(\bmk)\tilde u_j(\bmk)}
\int \frac{d^3k'}{(2\pi)^3}\ \tilde B_m(\bmk')
+i\abra{\tilde u_i^*(\bmk)\frac{\partial\tilde u_j(\bmk)}{\partial k_n}}
\int \frac{d^3k'}{(2\pi)^3}\ (ik'_n)\tilde B_m(\bmk') 
\right].
\end{align}
The two integrals inside the square brackets are equal to $B_m(\bmx=\bm 0)$ and $\Lambda_{nm}$, respectively.
Focusing on the term linear in $\bm\Lambda$, notice that we only need the real part of its pre-factor, because only the real part of $\tilde C_{ij}$ contributes to the EMF and all other tensorial coefficients in Equation (\ref{eqn:NCT}) are real.
Consequently we consider
\beq
\abra{\tilde u_i^*(\bmk)\frac{\partial\tilde u_j(\bmk)}{\partial k_n}}
\to
\frac{1}{2}\left(
\abra{\tilde u_i^*(\bmk)\frac{\partial\tilde u_j(\bmk)}{\partial k_n}}
+\abra{\tilde u_i(\bmk)\frac{\partial\tilde u_j^*(\bmk)}{\partial k_n}}
\right).
\eeq
The part that is symmetric in $i\leftrightarrow j$ in this tensor is
\begin{align}
\abra{\tilde u^*_i(\bmk)\frac{\partial\tilde u_j(\bmk)}{\partial k_n}}_{
\text{symmetric in }i,j}
=&\frac{1}{4}\left(
\abra{\tilde u_i^*(\bmk)\frac{\partial \tilde u_j(\bmk)}{\partial k_n}}
+\abra{\tilde u_i(\bmk)\frac{\partial \tilde u_j^*(\bmk)}{\partial k_n}}
+\abra{\tilde u_j^*(\bmk)\frac{\partial \tilde u_i(\bmk)}{\partial k_n}}
+\abra{\tilde u_j(\bmk)\frac{\partial \tilde u_i^*(\bmk)}{\partial k_n}}
\right)\notag\\
=&\frac{1}{2}\frac{\partial}{\partial k_n}\tilde K_{ij}.
\label{eqn:dKijdk}
\end{align}
On the other hand the anti-symmetric part is of the form $\epsilon_{ijl}T_{ln}$, where $T_{ln}$ is some real pseudotensor.
If the turbulence is homogeneous isotropic and nonhelical, the only brick to build $T_{ln}$ will be $k_i$, but it is then impossible to make it a pseudo-tensor.
Therefore the real, anti-symmetric part is zero.
This will not be the case if we consider the next order terms in the background turbulence which includes large-scale pseudo-vectors like $\Omega_i$ or $\del\times\bm U$.

Collecting only the terms linear in $\bm\Lambda$ we arrive at
\beq
\tilde T^{B,u}_{ij}
=-\frac{1}{2}\Lambda_{nm}k_m\frac{\partial\tilde K_{ij}}{\partial k_n}
+\text{terms linear in }\bmB.
\eeq
The magnetic contribution can be similarly worked out, and altogether we have
\beq
\tilde T^{B}_{ij}
=-\frac{1}{2}\Lambda_{nm}k_m\frac{\partial}{\partial k_n}\left(
\tilde K_{ij}+\tilde M_{ij}\right)
+\text{terms linear in }\bmB.
\eeq

\section{Anisotropy of velocity correlation tensor in rotating turbulence}
\label{appx:corfunc}
With isotropic forcing, a rigid rotation $\bm\Omega$, and a dynamically weak large-scale magnetic field, the most general form of $\tilde K_{ij}=\abra{\tilde u_i^*(\bmk)\tilde u_j(\bmk)}$ in a steady state is
\beq
\tilde K_{ij}=AP_{ij}+B\left(
\hat\Omega_{ij}+\hat k_{ij}\mu^2
-\hat k_{il}\hat\Omega_{lj}
-\hat\Omega_{il}\hat k_{lj}\right)
+iC\epsilon_{ijl}k_l,
\eeq
where $\mu=\hat{\bm k}\cdot\hat{\bm\Omega}$, and $A$, $B$ and $C$ are functions of $(k,\mu,\Omega)$ (not to be confused with the magnetic vector potential, the magnetic field, or the cross-correlation tensor in the main text).
Since they do not depend on the azimuthal direction $\phi$ in the $\bmk$-space, 
we will have $\tilde K_{ij}=\int \tilde K_{ij}\ d\phi/(2\pi)$, and thus
\begin{align}
&\tilde K_{11}=\tilde K_{22}
=\pi\left[(1+\mu^2)A+\frac{1}{2}\mu^2(1-\mu^2)B\right],\\
&\tilde K_{33}
=2\pi(1-\mu^2)(A+B-\mu^2B),\\
&K_{12}=-K_{21}
=2\pi i \mu C.
\end{align}
Solving for $A,B$ and $C$ we get
\begin{align}
&A=\frac{4(1-\mu^2)\tilde K_{11}-\mu^2\tilde K_{33}}{2\pi(1-\mu^2)(2+\mu^2)}
,\\
&B=\frac{-2(1-\mu^2)\tilde K_{11}+(1+\mu^2)\tilde K_{33}}{\pi(1-\mu^2)^2(2+\mu^2)}
,\\
&C=-\frac{i \tilde K_{12}}{2\pi\mu}.
\end{align}
These are measurable from simulations in order to quantify the anisotropy of $K_{ij}$.
To avoid the divergence in the denominators, in practice we will measure
\beq
A'=(1-\mu^2)A,\ B'=(1-\mu^2)^2B,\ C'=\mu C,
\eeq
so that
\begin{align}
&A=\left[
\frac{2}{3}P_0(\mu)-\frac{2}{3}P_2(\mu)
\right]^{-1} A',\\
&B=\left[
\frac{8}{15}P_0(\mu)-\frac{16}{21}P_2(\mu)+\frac{8}{35}P_4(\mu)
\right]^{-1} B',\\
&C=\left[P_1(\mu)\right]^{-1} C'.
\end{align}

The anisotropy of the normalized coefficients $A'$, $B'$, and $C'$ can be quantified through their Legendre decomposition, and we compute their Legendre coefficients as
\beq
a_l^I (k,\Omega)=\frac{2l+1}{2}
\int_{-1}^1I(k,\mu,\Omega)P_l(\mu)\ d\mu,
\eeq
where $I=A',B'$ or $C'$, so that
\beq
I=\sum_{l=0}^{\infty} a_l^I(k,\Omega)P_l(\mu).
\eeq

We performed hydrodynamical simulations using the $\texttt{Pencil Code}$ \citep{pc2021} with $256^3$ resolution and different Rossby numbers $\Ro=\urms\kf/(2\Omega)$.
The Reynolds numbers vary from $28$ to $67$ for the three runs.
The Legendre coefficients are computed up to the 10th order at wavenumber $k_l=13$, several wavenumbers beyond the forcing wavenumber $\kf=5$.
This is for the purpose of avoiding the dominant influence of the isotropic forcing at $\kf$, and correctly resolving the anisotropy in the inertial range.
Furthermore, the chosen $k_l$ is where the strongest anisotropy lies for the run with the highest rotation rate.
Snapshots of the azimuthally averaged correlation functions are computed and a further temporal average is taken to obtain $\tilde K_{ij}$.

\begin{figure}
\centering
\includegraphics[width=\columnwidth]{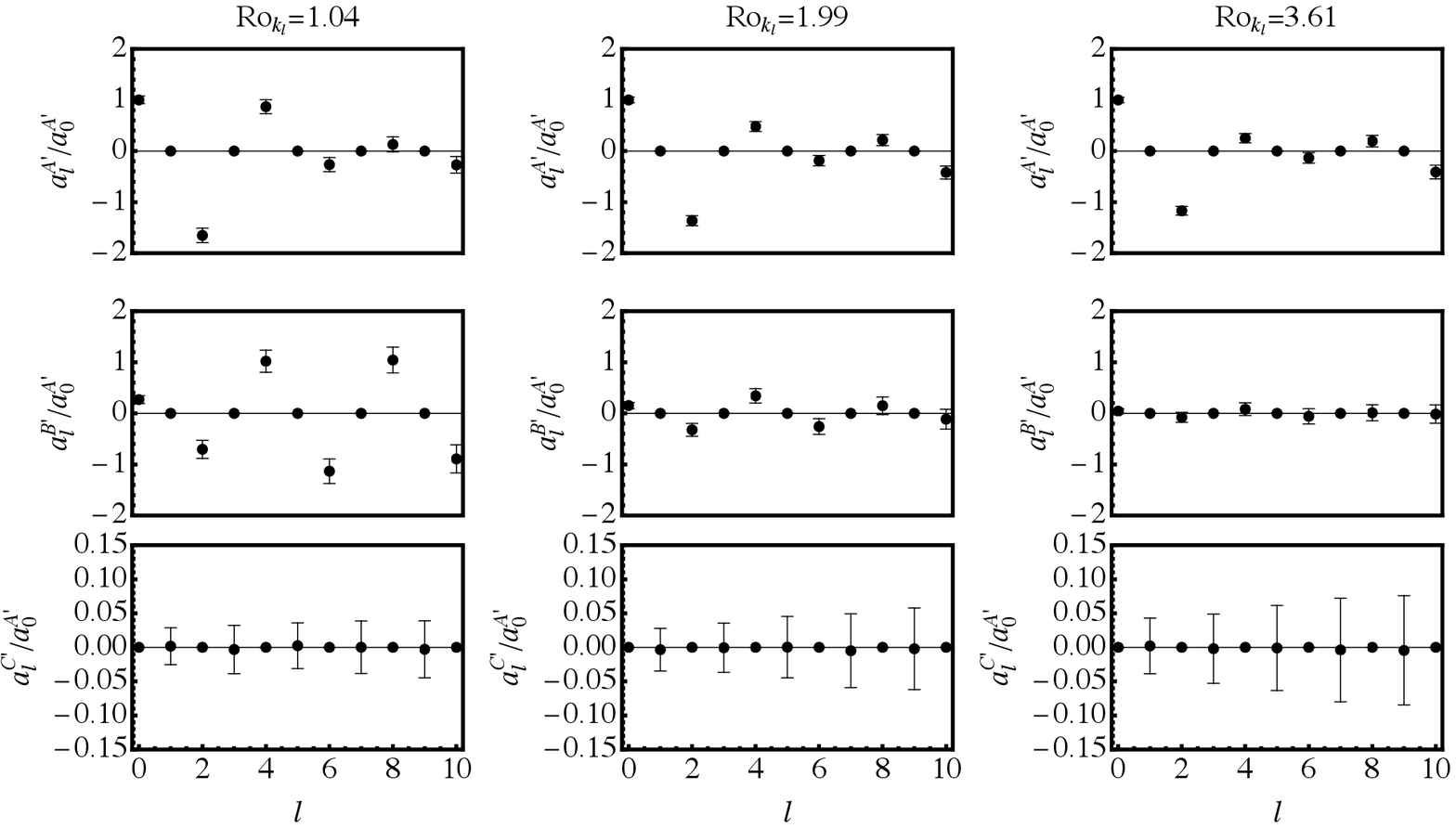}
\caption{Legendre coeffcients of the velocity correlation.}
\label{appx_fig:ABCal}
\end{figure}

In Figure \ref{appx_fig:ABCal}, each column shows the Legendre coefficients computed for $A'$, $B'$, and $C'$, for different Rossby numbers at $k_l$: $\Ro_{k_l}=(k_l/\kf)^{2/3}\Ro$.
For each column, the Legendre coefficients are normalized by $a_0^{A'}$.
The first and second rows reflect that the amplitudes of the anisotropic modes increase with decreasing Rossby number, and become dominant at $\Ro\lesssim2$.
Thus for $\Ro\gtrsim2$, $\tilde K_{ij}\propto P_{ij}$ remains a good approximation.

We do not explore numerically how the anisotropy of $K_{ij}$ will depend on $S$ here, but assume that isotropy remains a valid assumption with small shear parameters in correspondence to the pure rotation case, $\Sh\lesssim q/(2\Ro)=q/4$ where $q=-\ln\Omega/\ln r$.

\section{Derivation of the $n$th-order solution of (33)}
\label{appx:Cn}
From Equation (\ref{eqn:induction}), the $n$th increment is
\begin{align}
\tilde C^{(n)}-\tilde C^{(n-1)}
=&-Sk_2\mathcal{G}\partial_1(\tilde C^{(n-1)}-\tilde C^{(n-2)})\notag\\
=&(-Sk_2)^2\mathcal{G}\partial_1\left[
\mathcal{G}\partial_1\left(
\tilde C^{(n-2)}-\tilde C^{(n-3)}\right)\right]\notag\\
=&\cdots\notag\\
=&(-Sk_2)^n\mathcal{G}\partial_1\left[
\mathcal{G}\partial_1\left(\cdots
\mathcal{G}\partial_1 \tilde C^{(0)}\right)\right],\ 
\text{$n$ number of $\mathcal{G}\partial_1$'s.}
\end{align}
For the sake of clear notation we have used for the moment $\partial_1\equiv\partial/\partial k_1$ and omitted the tensor indices:
The four neighboring indices of each adjacent tensors will be contracted, e.g.,
$\left(\mathcal{G}\partial_1\mathcal{G}\right)_{abcd}=\mathcal{G}_{abef}\partial_1\mathcal{G}_{efcd}$ and so on.

Note that to compute the EMF we need to integrate $\tilde C_{ab}$ in the $\bmk$-space, allowing us to extract total derivatives which will not contribute to the EMF,
\begin{align}
\tilde C^{(n)}-\tilde C^{(n-1)}
=&(-Sk_2)^n\mathcal{G}\partial_1\left[
\mathcal{G}\partial_1\left(\cdots
\mathcal{G}\partial_1 \tilde C^{(0)}\right)\right]\notag\\
=&(-Sk_2)^n(-1)\partial_1\mathcal{G}\left[
\mathcal{G}\partial_1\left(\cdots
\mathcal{G}\partial_1 \tilde C^{(0)}\right)\right]+\text{total derivatives}\notag\\
=&(-Sk_2)^n(-1)^2\partial_1\left(\partial_1\mathcal{G}\mathcal{G}\right)
\left[
\mathcal{G}\partial_1\left(\cdots
\mathcal{G}\partial_1 \tilde C^{(0)}\right)\right]+\text{total derivatives}\notag\\
=&\cdots\notag\\
=&(Sk_2)^n\mathcal{H}^{(n)}\tilde C^{(0)}+\text{total derivatives},
\end{align}
where
\beq
\mathcal{H}^{(n)}=\partial_1\left[
\partial_1\left(\cdots\partial_1\mathcal{G}\mathcal{G}\right)
\mathcal{G}\right],\ 
\text{$n$ number of $\partial_1$'s, and $n$ number of $\mathcal{G}$'s},
\eeq
or equivalently
\beq
\mathcal{H}^{(1)}_{abcd}=\partial_1\mathcal{G}_{abcd},\ 
\mathcal{H}^{(n)}_{abcd}=\partial_1\left(\mathcal{H}^{(n-1)}_{abef}
\mathcal{G}_{efcd}\right).
\eeq
The $n$th-order solution is then
\begin{align}
\tilde C^{(n)}_{ab}=&\left[\delta_{ac}\delta_{bd}
+\sum_{m=1}^n (Sk_2)^m\mathcal{H}^{(m)}_{abcd}\right]
\tilde C^{(0)}_{cd}+\text{total derivatives}
\notag\\
=&\left[\tilde T^\Lambda_{cd}+\frac{1}{2}\Lambda_{nm}k_m
(\tilde K_{cd}+\tilde M_{cd})
\frac{\partial}{\partial k_n}\right]\left[
\mathcal{G}_{abcd}
+\sum_{m=1}^n (Sk_2)^m\mathcal{H}^{(m)}_{abef}
\mathcal{G}_{efcd}\right]
+\text{total derivatives}.
\end{align}
In the last step, we have additionally performed an integration by parts regarding the $\tilde T^B_{cd}$ term in $\tilde C^{(0)}_{cd}$.

\end{document}